\documentclass[aps,pra,superscriptaddress,twocolumn,amsmath,amssymb,floatfix,nofootinbib]{revtex4-2}

\pdfoutput=1
\usepackage{xcolor}
\definecolor{myblue}{rgb}{0.2,0.2,0.8}
\definecolor{myred}{rgb}{1,0.,0.3}
\definecolor{mygreen}{rgb}{0,0.57,0.0}
\usepackage{dcolumn}
\usepackage{bm}
\usepackage[colorlinks=true, citecolor=myblue, linkcolor=myred, urlcolor=myblue]{hyperref}
\usepackage{graphicx}
\usepackage{physics}
\usepackage{mathtools}
\usepackage{amsmath}
\usepackage{multirow}
\usepackage{dsfont}
\usepackage{siunitx}
\usepackage{layouts}
\usepackage{float}

\newcommand{\redr}{\textcolor{red}{r}}
\newcommand{\greeng}{\textcolor{mygreen}{g}}
\newcommand{\blueb}{\textcolor{blue}{b}}

\newcommand{\rwth}{Institute for Quantum Information, RWTH Aachen University, 52056 Aachen, Germany}
\newcommand{\fzj}{Peter Grünberg Institute, Theoretical Nanoelectronics, Forschungszentrum Jülich, 52425 Jülich, Germany}
\newcommand{\stuttgart}{Institute for Theoretical Physics III and Center for Integrated Quantum Science and Technology, University of Stuttgart, 70550 Stuttgart, Germany}

\begin{document}

\title{Multiqubit Rydberg Gates for Quantum Error Correction}

\author{David F. Locher} \email{david.locher@rwth-aachen.de}  \affiliation{\fzj}\affiliation{\rwth}
\author{Josias Old} \affiliation{\fzj}\affiliation{\rwth} 
\author{Katharina Brechtelsbauer} \affiliation{\stuttgart}
\author{Jakob Holschbach} \affiliation{\fzj}\affiliation{\rwth} 
\author{Hans Peter Büchler} \affiliation{\stuttgart}
\author{Sebastian Weber} \affiliation{\stuttgart}
\author{Markus Müller} \affiliation{\fzj}\affiliation{\rwth}

\date{May 4, 2026}

\begin{abstract}
Multiqubit gates that involve three or more qubits are usually thought to be of little significance for fault-tolerant quantum error correction because single gate faults can lead to errors of high Pauli weight. However, recent works have shown that multiqubit gates can be beneficial for measurement-free fault-tolerant quantum error correction and for fault-tolerant stabilizer readout in unrotated surface codes.
In this work, we investigate multiqubit Rydberg gates that are useful for fault-tolerant quantum error correction in single-species neutral-atom platforms and can be implemented with global laser pulses that do not individually address atomic sites. We develop an open-source Python package to generate analytical, few-parameter pulses that implement the desired gates while minimizing gate errors due to Rydberg-state decay. The tool also allows us to identify parameter-optimal pulses, characterized by a minimal parameter count for the pulse ansatz.
Measurement-free quantum error correction protocols require CCZ gates, which we analyze for atoms arranged in symmetric and asymmetric configurations. We investigate the performance of these schemes for various single-, two-, and three-qubit gate error rates, showing that break-even performance of measurement-free QEC is within reach of current hardware.
Moreover, we study Floquet quantum error correction protocols that comprise two-body stabilizer measurements. Those can be realized using global three-qubit gates, and we show that this can lead to a significant reduction in shuttling operations. Simulations with realistic circuit-level noise indicate that applying three-qubit gates for stabilizer measurements in Floquet codes can yield competitive logical qubit performance in experimentally relevant error regimes.
\end{abstract}

\maketitle

\iffalse
\setcounter{tocdepth}{2}
\begingroup
\hypersetup{linkcolor=black}
\tableofcontents
\endgroup
\fi

\section{Introduction}\label{sec:introduction}

Multiqubit gates that involve three or more qubits can be very valuable in quantum information processing. Although not required as part of a universal gate set, their availability can significantly decrease circuit depths~\cite{nielsenchuang2010quantum}.
Such potential resource savings apply in particular to fault-tolerant quantum computing protocols, where multiqubit gates occur, for example, in measurement-free quantum error correction (QEC)~\cite{heussen2024measurementfree, perlin2023fault, veroni2024optimized, brechtelsbauer2025measurementfree} and in measurement-free universal quantum computing~\cite{butt2024measurement, veroni2025universal} schemes.
Recently, it was established that multiqubit gates can also be applied for fault-tolerant stabilizer readout in unrotated surface codes~\cite{old2025faulttolerant, pecorari2025lowdepth}.

Several quantum computing platforms have demonstrated native implementations of multiqubit gates~\cite{molmer1999multipartite, kim2022highfidelity, levine2019parallel}.
In this work, we focus on neutral atoms temporarily excited to Rydberg states in order to execute two- and multiqubit gates~\cite{saffman_2010_quantum, saffman_2016_quantum, henriet2020quantum, morgado_2021_quantum}. This platform has seen multiple demonstrations of multiqubit gates~\cite{levine2019parallel, mcdonnel2022demonstration, evered2023highfidelity, cao2024multi}, and many more theory proposals for various types of multiqubit gates have been put forward~\cite{mueller2009mesoscopic, khazali2020fast, young2021asymmetric, jandura2022timeoptimal, pelegri2022high, dlaska2022quantum, kazemi2025multiqubit, pecorari2025lowdepth}.
Moreover, neutral-atom quantum processors are a leading platform for experimental demonstrations of elements from fault-tolerant quantum computing~\cite{bluvstein2024logical, reichardt2025faulttolerant, bedalov2024faulttolerant, rodriguez2025experimental, bluvstein2025architectural, rines2025demonstration}.

This paper investigates multiqubit gates implemented by global Rydberg laser pulses, i.e., laser beams that illuminate all atoms homogeneously.
We optimize gate pulses in the presence of Rydberg-state decay to identify gate implementations that minimize decay-induced errors. For this, we develop an optimization routine based on the \emph{chopped random basis} (CRAB) quantum control paradigm~\cite{caneva2011chopped, doria2011optimal, mueller2022one} that yields analytical pulse shapes described by a small number of parameters. This distinguishes the approach from \emph{gradient ascent pulse engineering} (GRAPE), a technique commonly used for the optimization of gate pulses~\cite{jandura2022timeoptimal, evered2023highfidelity, kazemi2025multiqubit}. Our Rydberg gate optimization software is available as an open-source Python package called \texttt{RydOpt}~\cite{rydopt}.

The types of multiqubit gates we investigate are useful for various fault-tolerant QEC protocols.
In particular, many measurement-free QEC protocols involve Toffoli-type gates on three or more qubits.
Interestingly, they often occur in a context where the control qubits are discarded after the gate. This relaxes constraints on the physical realization of these gates because one may add arbitrary single- and multi-qubit unitaries on those qubits just before the reset.
Another application that can profit from multiqubit gates is Floquet QEC~\cite{hastings2021dynamically, gidney2021faulttolerant, haah2022boundaries, gidney2022benchmarking, ustun2024single}, which requires repeated measurements of two-body stabilizers. We show that the mapping of these operators onto ancilla qubits can be achieved using global three-qubit gates, which in turn reduces circuit depths and simplifies shuttling schedules for neutral-atom implementations.

\begin{figure*}[t]
    \centering
    \includegraphics[width=\linewidth]{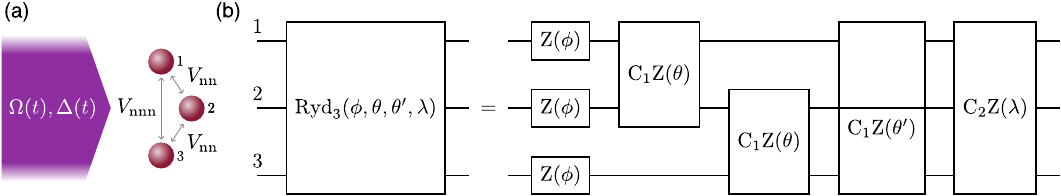}
    \caption{\textbf{Multiqubit Rydberg gates on three atoms.} (a) Three atoms arranged in an isosceles triangle are illuminated by a global Rydberg laser beam. This is not the most general arrangement of three atoms; however, it encompasses many geometries one might encounter in a lattice (e.g.~equilateral triangle, right triangle, or straight line). The interaction between the first atom and the third one may be different from the interaction between the second atom and each of the other ones. This allows, in principle, for the realization of multiqubit gates that are not symmetric with respect to the exchange of any two qubits. (b) A parameterization of all multiqubit gates that the global Rydberg pulse can realize. The constituent gates are defined as $\mathrm{C}_{n}\mathrm{Z}(\alpha) := \mathrm{diag}(1, ..., 1, e^{i\alpha})$ on $n+1$ qubits, and $\mathrm{Z}(\alpha) := \mathrm{C}_{0}\mathrm{Z}(\alpha) = \mathrm{diag}(1, e^{i\alpha})$.}
    \label{fig:Rydberg_gate_circuit_3}
\end{figure*}

The paper is structured as follows. Section~\ref{sec:multiqubit_gates} is a technical chapter that describes the physics of multiqubit Rydberg gates and presents our optimization procedure. Moreover, it demonstrates the optimization routine's capabilities with the examples of CZ and CCZ gates implemented with pulses that minimize either the gate duration (time-optimal), the time spent in Rydberg states (Rydberg time-optimal), or the pulse's total parameter count (parameter-optimal).
Section~\ref{sec:mf_qec} then discusses a first application that benefits from the availability of multiqubit gates, which is measurement-free fault-tolerant QEC. The respective protocols require CCZ gates, which we analyze for non-symmetric atomic geometries. We investigate the gate fidelities required for CCZ gates to be beneficial for measurement-free QEC, showing that break-even performance of measurement-free QEC protocols could be demonstrated on current hardware.
Section~\ref{sec:floquet} presents a Floquet quantum error correction scheme that can benefit from global three-qubit gates for two-body stabilizer readout due to efficient shuttling schedules and biased noise characteristics of Rydberg gates.
Finally, Sec.~\ref{sec:conclusion} concludes the paper.

\section{Multiqubit Rydberg Gates}\label{sec:multiqubit_gates}

    This section provides a brief summary of the physical details underlying two- and multiqubit Rydberg gates, and outlines which types of gates are realizable using only global laser pulses. We then carefully describe our pulse optimization technique, which yields pulses that implement high-fidelity gates and can be characterized analytically with just a few parameters. Finally, we discuss time-optimal and Rydberg time-optimal CZ and CCZ gates.

\subsection{Physical details}

    State-of-the-art neutral-atom quantum computing setups~\cite{henriet2020quantum, morgado_2021_quantum} consider qubits encoded in two low-lying electronic states of the atoms. Temporarily exciting atoms to a Rydberg state $\ket{r}$ allows one to realize two- and multiqubit gates due to strong van der Waals interactions between highly excited atoms~\cite{jaksch2000fast, saffman_2010_quantum, saffman_2016_quantum}.
    Transitions between the atomic states $\ket{1}$ and $\ket{r}$ are driven by a laser with detuning $\Delta(t)$ and complex Rabi frequency $\Omega(t) = |\Omega(t)|e^{i \xi(t)}$.
    A laser illuminating $N$ atoms homogeneously realizes the Hamiltonian
    \begin{equation} \label{eq:Hamiltonian}
        \begin{split}
            H(t) = & \; \hbar \sum_{i=1}^{N} \frac{1}{2} \big( \Omega(t) \ket{r_i}\!\bra{1_i} + \mathrm{h.c.} \big) - \Delta(t) \ket{r_i}\!\bra{r_i} \\
            & + \sum_{i, j<i} V_{ij} \ket{r_i r_j}\!\bra{r_i r_j} ,
        \end{split}
    \end{equation}
    where $V_{ij}$ denotes the van der Waals interaction between the atoms $i$ and $j$, whose precise value depends on the atomic details. In general, it decays with increasing distance $d$ between the two atoms as $1/d^6$.
    Two-photon Rydberg excitations can also be described by Eq.~\eqref{eq:Hamiltonian}; $\Omega(t)$ and $\Delta(t)$ then correspond to the effective two-photon Rabi frequency and two-photon detuning, respectively.

    Two- or multiqubit quantum gates are implemented by laser pulses, i.e.,~variations in time of the Rabi frequency and possibly also the detuning. The pulse temporarily excites atoms to Rydberg states and eventually brings all populations back to the computational space.
    For simplicity, we consider pulses with a fixed Rabi frequency amplitude $|\Omega(t)| = \Omega_0$ for the entire pulse duration $T$. This is not a severe restriction, as we study fast, strongly non-adiabatic pulses in this paper. The quantity $\Omega_0$ sets a natural scale for quantifying times and frequencies.
    In an experiment, the laser power and, thus, the Rabi frequency amplitude would be ramped up and down smoothly at the beginning and the end of the pulse, respectively. Pulse optimizations can just as well be performed for time-dependent Rabi frequency profiles, which our software also supports.
    One may choose to set the detuning to zero and vary only the laser phase in time. Alternatively, one can fix the laser phase to be zero and implement a time-dependent detuning. This is possible because a pulse A described by $\xi_{\mathrm{A}}(t)$ and $\Delta_{\mathrm{A}}(t) = 0$ can be shown to implement the same gate as a pulse B with $\xi_{\mathrm{B}}(t) = 0$ and $\Delta_{\mathrm{B}}(t) = -\frac{d\xi_{\mathrm{A}}(t)}{dt}$.
    This can be derived from a time-dependent unitary transformation of the Hamiltonian~\eqref{eq:Hamiltonian}.

    A laser pulse described by the Hamiltonian in Eq.~\eqref{eq:Hamiltonian} does not affect the qubit states $\ket{0_i}$. Therefore, the Hamiltonian exhibits a block-diagonal structure with one block for each computational basis state $\ket{\{0,1\}^N}$. The types of gates a Rydberg pulse can realize are thus gates that imprint phases onto each computational basis state.
    Depending on the atomic geometry and hence the interactions present between the atoms, some of the Hamiltonian's blocks are equivalent and thus pick up the same phase.
    This also implies that an atomic geometry that is not symmetric under the exchange of any two atoms allows one to implement gates that are not completely symmetric, even though the very same global laser field illuminates all atoms, with no individual addressing~\cite{stein2025multitarget, kazemi2025multiqubit}.
    Figure~\ref{fig:Rydberg_gate_circuit_3} shows a parameterization for all possible multiqubit Rydberg gates on three atoms arranged in an isosceles triangle. We define an $n$-qubit-controlled $\mathrm{Z}( \alpha )$ gate as
    \begin{equation}
        \mathrm{C}_{n}\mathrm{Z}( \alpha ) = \mathrm{diag}\left(1, ..., 1, e^{i\alpha} \right) , \quad \mathrm{dim}\left( \mathrm{C}_{n}\mathrm{Z} \right) = 2^{n+1} .
    \end{equation}
    The asymmetry of the atomic arrangement is reflected in two different phases $\theta$ and $\theta'$ in the circuit. For a symmetric geometry with $V_{\mathrm{nn}} = V_{\mathrm{nnn}}$, one finds that $\theta' = \theta$.
    Single-qubit phases $\phi$ imprinted as a part of a Rydberg gate $\mathrm{Ryd}_3(\phi, \theta, \theta', \lambda)$ can be modified by a global single-qubit gate pulse. Therefore, one usually allows a Rydberg pulse to implement an arbitrary value of $\phi$ that can be easily cancelled to realize the desired three-qubit gate $G_3$:
    \begin{equation}
        G_3(\theta, \theta', \lambda) = \mathrm{Ryd}_3(\phi, \theta, \theta', \lambda) \ \mathrm{Z}(-\phi)^{\otimes 3} .
    \end{equation}

\subsection{Pulse engineering}

    A theoretically simple yet powerful pulse ansatz is the expansion of the laser phase or the laser detuning in trigonometric functions with adjustable frequencies. This technique is known under the name \emph{chopped random basis} (CRAB)~\cite{caneva2011chopped, doria2011optimal, mueller2022one}.
    An advantage of this ansatz is that the pulse complexity can be increased systematically by adding terms with higher frequencies, thereby introducing more pulse parameters. Moreover, the resulting pulses are naturally smooth and can be described analytically. Also, the pulses require only a limited bandwidth.
    As noted above, we may choose to vary only the laser phase in time.
    For Rabi frequency $\Omega(t) = \Omega_0 e^{i \xi(t)}$ and detuning $\Delta(t) = \Delta_0$, such a pulse may be written as follows:
    \begin{equation} \label{eq:ansatz_sin_cos}
    \begin{split}
        \xi(t) = & \sum_{n=1}^{K} \alpha_n \sin\left(\frac{2 \pi}{T} n \left(1 + \tfrac{1}{2} \tanh(A_n)\right)\left(t - \tfrac{T}{2}\right)\right) \\
        & + \beta_n \cos\left(\frac{2 \pi}{T} n \left(1 + \tfrac{1}{2} \tanh(B_n)\right)\left(t - \tfrac{T}{2}\right)\right),
    \end{split}
    \end{equation}
    with $t \in [0, T]$. The parameters describing the pulse are
    \begin{equation}
        \big(T, \Delta_0, A_1, \alpha_1, B_1, \beta_1, ..., A_K, \alpha_K, B_K, \beta_K \big) .
    \end{equation}
    We see that this ansatz is essentially a truncated Fourier series with coefficients $\alpha_n$ and $\beta_n$, where the frequencies of all terms can be adjusted around the respective principal harmonics by the parameters $A_n$ and $B_n$. For $A_n = B_n = 0$, we retrieve the principal harmonics of a standard Fourier series.
    Note that the gate duration $T$ is one of the optimization parameters, i.e., it is not fixed beforehand, as in some other pulse optimization methods.

    A more restrictive ansatz is the expansion of the laser phase in sine terms only, resulting in pulse profiles that are antisymmetric:
    \begin{equation} \label{eq:ansatz_sin}
        \xi(t) = \sum_{n=1}^{K} \alpha_n \sin\left(\frac{2 \pi}{T} n \left(1 + \tfrac{1}{2} \tanh(A_n)\right)\left(t - \tfrac{T}{2}\right)\right)
    \end{equation}
    with $t \in [0, T]$, $\Delta(t) = \Delta_0$, and pulse parameters
    \begin{equation}
        \big(T, \Delta_0, A_1, \alpha_1, ..., A_K, \alpha_K \big) .
    \end{equation}
    We will see later that for small numbers of pulse parameters ($\approx 10$) this ansatz yields much better results than the more general ansatz specified in Eq.~\eqref{eq:ansatz_sin_cos}. Considering small numbers of pulse parameters is not only helpful in order to make the numerical gate optimization more efficient. Also in experiments, it is convenient to calibrate just a few parameters that specify the gate. Therefore, throughout the paper, we mainly consider ansatz~\eqref{eq:ansatz_sin}.
    
    Note that the constant detuning $\Delta_0$ is equivalent to a linear contribution $-\Delta_0 t$ to the phase profile $\xi(t)$ and, hence, could be replaced accordingly.
    It is also possible to split this term into two summands $\Delta_0 = \Delta_0' + \Delta_0''$. One part can be implemented as a constant detuning, $\Delta(t) = \Delta_0'$, while the other part is realized as a linear contribution to the phase, $-\Delta_0'' t$. By choosing $\Delta_0''$ appropriately, one could achieve that the pulse profile given in Eq.~\eqref{eq:ansatz_sin} begins and ends with $\tfrac{d\xi}{dt}\!\big|_{0}=\tfrac{d\xi}{dt}\!\big|_{T}=0$.

    In order to implement a desired gate, one must find an appropriate set of pulse parameters given a specific pulse ansatz.
    We implement the gate optimization in Python using the package \texttt{JAX}~\cite{jax}. Our code is publicly available on GitHub in the form of a Python package that we call \texttt{RydOpt}~\cite{rydopt}. The time evolution of an arbitrary computational state can be determined by calculating the dynamics of each computational basis state individually, since they all belong to distinct blocks of the Hamiltonian. We use the Python package \texttt{Diffrax}~\cite{diffrax} to solve the Schrödinger equation for each subsystem numerically. Given a particular target gate $U_{\mathrm{targ}}$, we perform gradient ascent (adam) using a generalized Bell state fidelity as a cost function
    \begin{equation} \label{eq:costfunction}
        F = |\! \bra{+}^{\otimes N} U_{\mathrm{targ}}^{\dagger} U(T) \ket{+}^{\otimes N}\!|^2 ,
    \end{equation}
    where $U(t)$ is the unitary evolution up to time $t$ generated by the pulse.
    Considering the example of three atoms, as shown in Fig.~\ref{fig:Rydberg_gate_circuit_3}, we specify a target gate by the phases $\theta$, $\theta'$, and $\lambda$, while allowing for arbitrary single-qubit rotations $\mathrm{Z}(\phi)$ on the qubits.
    One might encounter applications where the gate phase $\theta'$ does not need to be fixed precisely. This can happen, e.g.,~because two qubits are discarded after the gate, as we will see in Sec.~\ref{sec:mf_qec}. In such cases, the gate optimization routines can also allow for arbitrary values of $\theta'$.

    We now briefly comment on some differences of our pulse optimization scheme and the method \emph{gradient ascent pulse engineering} (GRAPE), used in many recent works on two- and multiqubit Rydberg gates~\cite{jandura2022timeoptimal, evered2023highfidelity, jandura2023optimizing, kazemi2025multiqubit}. GRAPE fixes a gate time $T$ and implements a pulse as a piecewise constant function, where the number of pieces is chosen to be large to achieve quasi-smooth functions. The approach allows one to search for arbitrary functions on the given time interval $[0,T]$. However, the number of optimization parameters is large, Ref.~\cite{jandura2022timeoptimal}, for example, employs between $99$ and $399$ parameters, and the resulting pulses that implement the gates can, in general, not be described analytically.
    Our approach, on the other hand, has the advantage that, by construction, the pulse functions are smooth and can be described analytically. Moreover, the pulse complexity, i.e.~the number of parameters required to describe the pulse, can be increased systematically by including more terms in the series. We find that typically a small number of optimization parameters is sufficient to design nearly time-optimal or Rydberg time-optimal multiqubit gates.

    \begin{figure}[t]
        \centering
        \includegraphics[width=\linewidth]{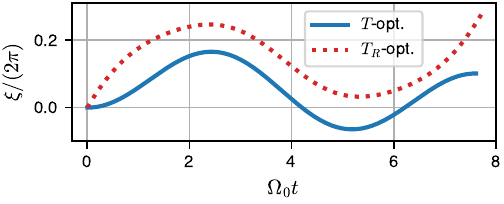}
        \caption{\textbf{Time-optimal and Rydberg time-optimal $\bm{\mathrm{CZ}}$ gate pulses in the perfect Rydberg blockade regime.} The time-optimal pulse~\cite{jandura2022timeoptimal, pagano2022error} requires only 4 parameters using the antisymmetric ansatz~\eqref{eq:ansatz_sin} ($\Omega_0 T = 7.611, \, \Omega_0 T_R = 2.958$). The Rydberg time $T_R$ can be reduced marginally when the pulse optimization is performed in the presence of decay from Rydberg states, which we model by adding a non-Hermitian term to the Hamiltonian (see Eq.~\eqref{eq:H_decay}). A 6-parameter pulse with ansatz~\eqref{eq:ansatz_sin} yields $\Omega_0 T = 7.725, \, \Omega_0 T_R = 2.936$. All pulse parameters are provided in Appendix~\ref{app:pulse_parameters}, Table~\ref{tab:timeoptimal_CZ}. Here and in all following plots of pulse profiles, a constant detuning $\Delta_0$ is translated into a linear contribution $-\Delta_0 t$ to the phase profile $\xi(t)$, and a constant is added such that $\xi(0)=0$.}
        \label{fig:CZ_pulses}
    \end{figure}

    \begin{figure}[t]
        \centering
        \includegraphics[width=\linewidth]{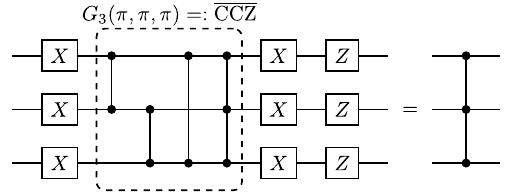}
        \caption{\textbf{Fast implementation of a Rydberg CCZ gate.} A CCZ gate can be realized by physically applying the gate \mbox{$G_{3}(\pi,\pi,\pi)$} $=: \overline{\mathrm{CCZ}}$ preceded and succeeded by global single-qubit rotations. As described in Ref.~\cite{evered2023highfidelity}, the $\overline{\mathrm{CCZ}}$ gate can be implemented with a faster global Rydberg pulse than the traditional CCZ gate. While the minimal pulse duration of the former gate is $\Omega_0 T = 10.8$, the latter one takes at least a time $\Omega_0 T = 16.4$ in the perfect blockade regime.}
        \label{fig:circuit_CCZ_fast}
    \end{figure}

\subsection{Time-optimal and Rydberg time-optimal CZ and CCZ gates}

    To demonstrate the capabilities of our pulse optimization routine, this subsection analyzes two-qubit CZ and three-qubit CCZ gates in the perfect blockade regime. We reproduce optimal gate times discussed in Refs.~\cite{pagano2022error, jandura2022timeoptimal, evered2023highfidelity}. Additionally, we investigate optimal Rydberg times for CCZ gates, and we study how gate times and Rydberg times depend on the pulse complexity.

    Starting with the analysis of the two-qubit CZ gate, Refs.~\cite{jandura2022timeoptimal, pagano2022error} have shown that the minimal duration of a pulse realizing this gate in the perfect blockade regime, i.e.~$V_{12}/(\hbar \Omega_0) = \infty$, is $\Omega_0 T= 7.61$.
    The time-optimal CZ gate can be implemented with a pulse described by just four parameters\footnote{Our ansatz does not support fewer than four parameters. However, almost time-optimal CZ gates can be realized with just three parameters. For example, fixing $\Delta_0=0$, one obtains a CZ gate that exhibits a pulse duration $\Omega_0 T = 7.643$. Alternatively, fixing $A_1=\infty$, one finds a CZ gate pulse with $\Omega_0 T = 7.612$.} using ansatz~\eqref{eq:ansatz_sin_cos}, or equivalently ansatz~\eqref{eq:ansatz_sin}. We show the pulse profile in Fig.~\ref{fig:CZ_pulses} and provide the pulse parameters in Appendix~\ref{app:pulse_parameters}, Table~\ref{tab:timeoptimal_CZ}.

    \begin{figure*}[t]
        \centering
        \includegraphics[width=\linewidth]{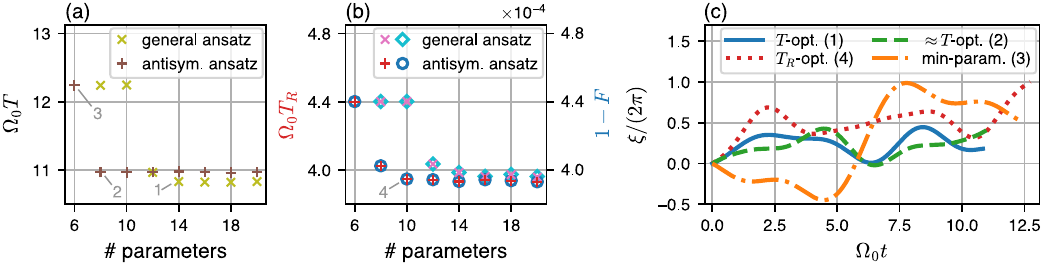}
        \caption{\textbf{Time-optimal and $\bm{T_R}$-optimal $\bm{\overline{\mathrm{CCZ}}}$ gates in the perfect Rydberg blockade regime.} For an increasing number of pulse parameters, and thus increasing pulse complexity, we optimize $\overline{\mathrm{CCZ}}$ gate pulses using the general ansatz~\eqref{eq:ansatz_sin_cos} and the antisymmetric ansatz~\eqref{eq:ansatz_sin}. Panel (a) shows the minimal pulse durations $\Omega_0 T$ found among $4\times10^4$ optimization runs for each data point. Every such pulse realizes the target gate with infidelity $<\!10^{-7}$ in the \emph{absence} of Rydberg-state decay. The time-optimal pulse~\cite{evered2023highfidelity}, marked with a grey number \textsf{1} in the plot, requires 14 parameters (\textsf{1}: $\Omega_0 T = 10.83, \, \Omega_0 T_R = 4.91$). A nearly time-optimal gate can be realized with a pulse described by 8 parameters (\textsf{2}: $\Omega_0 T = 10.97, \, \Omega_0 T_R = 4.18$). The minimal number of pulse parameters required to realize the gate is 6 (\textsf{3}: $\Omega_0 T = 12.24, \, \Omega_0 T_R = 4.40$). In panel (b), pulses are optimized in the presence of Rydberg-state decay with rate $\gamma/\Omega_0 = 10^{-4}$. Again, each data point corresponds to the best out of $4\times10^4$ optimization runs. In order to achieve the smallest possible gate error, the Rydberg time $T_R$ is minimized. We calculate both quantities separately, confirming $1-F=\gamma T_R$ for $\gamma T_R \ll 1$. A 10-parameter pulse can reduce $T_R$ considerably compared to the time-optimal gates (\textsf{4}: $\Omega_0 T = 12.73, \, \Omega_0 T_R = 3.95$). Note that a smaller gate duration does not imply smaller Rydberg times $T_R$. Panel (c) shows the pulse profiles for the gates \textsf{1}--\textsf{4} discussed in panels (a, b). The respective pulse parameters are provided in Appendix~\ref{app:pulse_parameters}, Table~\ref{tab:timeoptimal_CCZ}.}
        \label{fig:CCZ_combined}
    \end{figure*}

    A very important quantity to consider is the average time spent in Rydberg states during the execution of the gate,
    \begin{equation} \label{eq:TR}
        T_R = \int_0^T \sum_{i=1}^{N} \bra{+}^{\otimes N}U(t)^{\dagger} \ketbra{r_i}  U(t)\ket{+}^{\otimes N} dt ,
    \end{equation}
    because leakage processes from Rydberg states, such as radiative decay, are a dominant cause of gate errors. In fact, for two-qubit CZ gates with fidelity $\gtrsim 0.999$, Rydberg decay is projected to be the principal error source~\cite{tsai2025benchmarking}.
    To minimize such errors, it is thus reasonable to search for pulses that minimize the Rydberg time $T_R$ while implementing the desired gate. To do so, one can simply add a decay term
    \begin{equation} \label{eq:H_decay}
        H_{\mathrm{decay}} = \sum_{j=1}^{N} - \frac{i}{2} \hbar \gamma \ketbra{r_j}
    \end{equation}
    to the Hamiltonian~\eqref{eq:Hamiltonian} describing the system dynamics. Since this contribution to the total Hamiltonian $H$ is non-Hermitian, the norm of a state $\ket{\psi}$ being evolved in time under the action of $H$ is not conserved. The term thus describes decay with rate $\gamma$ from the Rydberg state into additional states that are not part of the Hilbert space described by $H$. Alternatively, one may interpret a decay event as a loss of atoms in our description. Unlike a Lindblad evolution, the approach does not allow one to model decay into states that evolve coherently under $H$. We are interested in the regime of small errors, in which this method provides a sufficiently accurate description. Moreover, it is computationally less demanding than an open-system simulation.
    An analogous non-Hermitian approach can be employed to model intermediate-state decay during two-photon Rydberg excitations. It is also supported by \texttt{RydOpt} and can capture, for example, different decay errors depending on the sign of the intermediate-state detuning, as discussed in Ref.~\cite{evered2023highfidelity}.

    Performing the pulse optimization in the presence of decay, we find that a $T_R$-optimal pulse implementing the CZ gate is described by six parameters and yields an improvement in $T_R$ of less than $1\%$ compared to the time-optimal pulse. This is hardly significant, in agreement with what has been found in Ref.~\cite{jandura2022timeoptimal}. Nevertheless, we plot the pulse in Fig.~\ref{fig:CZ_pulses} as well.
    
    We now go on to the CCZ gate realized with a global Rydberg laser beam. This gate has been analyzed in Ref.~\cite{jandura2022timeoptimal}, which discusses that in the perfect blockade regime, i.e.~$V_{ij}/(\hbar \Omega_0) = \infty \; \forall \, i,j$, the minimal duration of a pulse implementing the gate is $\Omega_0 T= 16.4$.
    It was found later and discussed in Ref.~\cite{evered2023highfidelity} that the gate duration can be further reduced by making use of the identity sketched in Fig.~\ref{fig:circuit_CCZ_fast}. To realize a CCZ gate, one can physically apply the global Rydberg gate $G_3(\pi, \pi, \pi) =: \overline{\mathrm{CCZ}}$ preceded and succeeded by global single-qubit gates. It turns out that the $\overline{\mathrm{CCZ}}$ gate can be implemented with a faster Rydberg laser pulse than the standard CCZ gate. Concretely, Ref.~\cite{evered2023highfidelity} finds the duration of the time-optimal $\overline{\mathrm{CCZ}}$ gate to be $\Omega_0 T = 10.8$.

    \begin{figure*}[t]
        \centering
        \includegraphics[width=\linewidth]{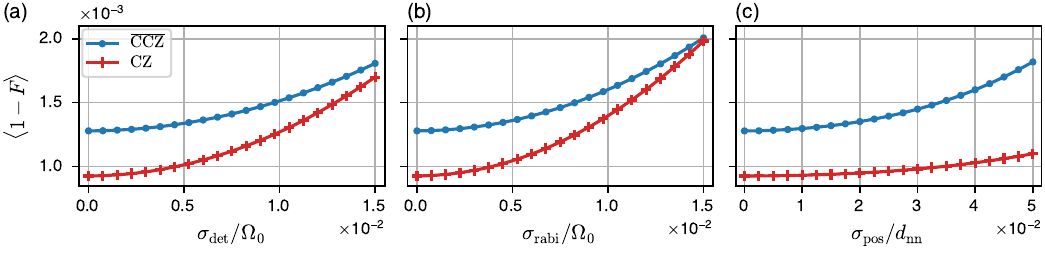}
        \caption{\textbf{Gate errors due to shot-to-shot fluctuations of the laser detuning (a), Rabi frequency amplitude (b), and atomic positions (c).} We consider Gaussian fluctuations with standard deviations $\sigma_{\mathrm{det}}$, $\sigma_{\mathrm{rabi}}$, and $\sigma_{\mathrm{pos}}$, respectively, to determine the expected gate infidelities $\langle 1-F \rangle$ of two-qubit CZ and three-qubit $\overline{\mathrm{CCZ}}$ gates. We employ a four-parameter CZ gate pulse and an eight-parameter $\overline{\mathrm{CCZ}}$ gate pulse using the antisymmetric ansatz~\eqref{eq:ansatz_sin}. Both pulses are optmized for interatomic interaction strengths $V_{\mathrm{nn}}/(\hbar \Omega_0)=32$. All panels include Rydberg-state decay with rate $\gamma/\Omega_0 = 3.18 \times 10^{-4}$.}
        \label{fig:gate_errors}
    \end{figure*}

    We analyze the $\overline{\mathrm{CCZ}}$ gate using the pulse ansätze specified in Eqs.~\eqref{eq:ansatz_sin_cos} and \eqref{eq:ansatz_sin}. The results are summarized in Fig.~\ref{fig:CCZ_combined}. Panel (a) investigates the time-optimal gate. For various numbers of pulse parameters, we perform $4 \times 10^4$ optimization runs and plot the gate time $\Omega_0 T$ of the fastest run that reaches an infidelity $<10^{-7}$.
    The pulse optimization is likely to converge to suboptimal solutions; therefore, running multiple independent optimization attempts increases the chance of finding a solution that is close to the global optimum.
    We find that a gate duration $\Omega_0 T = 10.83$ requires 14 pulse parameters using the general pulse ansatz. However, it turns out that a nearly time-optimal gate with $\Omega_0 T = 10.97$ can be realized with a pulse that is described by just eight parameters using the antisymmetric ansatz. These results are consistent with what has been found in Ref.~\cite{evered2023highfidelity}.
    The gate can even be implemented with a pulse described by only six parameters. This closely matches the number of constraints that must be satisfied to implement the desired gate: In the perfect blockade case, there are three distinct subsystems (representatives of which are the states $\ket{100}$, $\ket{110}$, and $\ket{111}$, respectively) whose populations must return to the computational space at the end of the gate. These subsystems have to acquire appropriate phases, up to a global single-qubit rotation, making up five constraints in total. Our ansatz supports only even numbers of parameters; thus, our minimal-parameter gate employs a six-parameter pulse. One might refer to a gate pulse that realizes a desired gate with minimal parameter count as a ``parameter-optimal'' pulse.
    
    An important observation is that a small gate time $\Omega_0 T$ does not imply a small Rydberg time $\Omega_0 T_R$. For example, among all pulses from panel (a), the pulse labeled~\textsf{2} exhibits the smallest Rydberg time with $\Omega_0 T_R=4.18$, although pulse~\textsf{1} is shorter (see e.g.~Appendix~\ref{app:pulse_parameters}, Table~\ref{tab:timeoptimal_CCZ}).
    To implement the gate with a pulse that minimizes $T_R$, we perform the pulse optimization in the presence of the decay term~\eqref{eq:H_decay} with $\gamma/\Omega_0 = 10^{-4}$ for various numbers of pulse parameters. Figure~\ref{fig:CCZ_combined}(b) shows Rydberg times $T_R$ and gate infidelities corresponding to the best pulses found among $4\times 10^4$ optimization attempts for each data point. As explained in Ref.~\cite{jandura2022timeoptimal}, a pulse that implements the gate perfectly in the absence of decay exhibits an infidelity $1-F=\gamma T_R$ if decay is present and $\gamma T_R \ll 1$. We find that the Rydberg time $T_R$ can be improved considerably compared to the time-optimal pulses. Figure~\ref{fig:CCZ_combined}(c) finally shows the pulse profiles for the (nearly) time-optimal gates, the minimal-parameter gate, as well as the $T_R$-optimal gate marked in the previous panels.
    In Appendix~\ref{app:pulse_parameters} we provide the pulse parameters in Table~\ref{tab:timeoptimal_CCZ}. Moreover, in Appendix~\ref{app:more_multiqubit_gates} we discuss time-optimal and $T_R$-optimal pulses implementing four-qubit $\overline{\mathrm{CCCZ}}$ gates, adding insights to what has been discussed in Ref.~\cite{evered2023highfidelity}.

\subsection{Errors on CZ and CCZ gates} \label{sec:errors_cz_ccz}

    Although \texttt{RydOpt} is primarily aimed at \emph{optimizing} gate pulses, one may use its routines to \emph{analyze} the performance of a gate in the presence of errors that extend beyond only Rydberg-state decay. Providing a microscopic error budget for multiqubit gates is beyond the scope of the present work. Still, in this subsection, we perform an approximate analysis of how multiple error channels affect the expected performance of three-qubit $\overline{\mathrm{CCZ}}$ gates compared to two-qubit CZ gates.
    
    Tsai \emph{et al.}~\cite{tsai2025benchmarking} perform detailed experimental and theoretical analyses of two-qubit gate errors. As mentioned in the previous subsection, they find that the fidelity of two-qubit gates is fundamentally limited by Rydberg-state decay. Laser frequency noise, intensity noise, and atomic motion are identified as the next-most important sources of error; however, the latter two are considerably less dominant than Rydberg decay. They remark that techniques such as cavity filtering can also reduce laser frequency noise to strongly subdominant levels.
    In our simulations, we incorporate Rydberg-state decay by adding a non-Hermitian term to the system Hamiltonian, as done for gate optimization. We model the other three types of errors as shot-to-shot fluctuations of the laser detuning $\Delta$, the Rabi frequency amplitude $|\Omega|$, and the atomic positions, the latter of which translate into fluctuations of atomic interaction strengths $V_{ij}$.
    Modeling those errors as shot-to-shot fluctuations is, of course, a simplistic approach that does not capture all physical details. A more exhaustive, microscopic analysis could, e.g., incorporate fluctuations during gate pulses. Also, laser-intensity inhomogeneities on the atoms can be modeled using \texttt{RydOpt}. Nevertheless, we compare the infidelities of three-qubit gates and two-qubit gates based on a simplified error model in order to obtain an estimate of achievable three-qubit gate fidelities.
    
    Concretely, we investigate CZ and $\overline{\mathrm{CCZ}}$ gate pulses, described by four and eight parameters, respectively, optimized for symmetric interaction strengths $V_{\mathrm{nn}} = 32\hbar \Omega_0$.
    This choice is based on realistic parameters discussed in Ref.~\cite{pagano2022error}. The Rydberg state $\ket{r} \! = \! \ket{60^3S_1, m_J\!=\!1}$ of $^{88}\mathrm{Sr}$ exhibits a van der Waals coefficient $C_6/h = \SI{-154}{\giga\hertz\,\micro\metre^6}$. With a Rydberg Rabi frequency $\Omega_0 = 2 \pi \times \SI{10}{\mega\hertz}$ and an interatomic spacing of $d_{\mathrm{nn}}=\SI{2.8}{\micro\metre}$ we obtain $V_{\mathrm{nn}}/(\hbar \Omega_0) = 32$ for the nearest-neighbour interaction strength. The Rydberg decay rate is estimated at $\gamma = \SI{20}{\kilo \hertz}$, corresponding to a ratio $\gamma/\Omega_0\approx3.18\times10^{-4}$. All these values are similar to those reported in the experimental demonstrations of Refs.~\cite{tsai2025benchmarking, evered2023highfidelity}.
    
    The three panels of Fig.~\ref{fig:gate_errors} show the gate infidelities resulting from shot-to-shot fluctuations of the laser detuning, the Rabi frequency amplitude, and the atomic positions, each including Rydberg-state decay. All fluctuations are modeled to be Gaussian, characterized by standard deviations $\sigma_{\mathrm{det}}$, $\sigma_{\mathrm{rabi}}$, and $\sigma_{\mathrm{pos}}$, respectively. In the presence of only Rydberg-state decay, the CZ-gate infidelity is estimated to be slightly below $10^{-3}$. The $\overline{\mathrm{CCZ}}$ gate exhibits an error of $1.28\times 10^{-3}$ considering only Rydberg decay, due to the slightly larger Rydberg time $\Omega_0 T_R$. Interestingly, shot-to-shot fluctuations of the laser detuning and Rabi frequency amplitude have a stronger influence on the CZ-gate fidelity than on the $\overline{\mathrm{CCZ}}$-gate fidelity. Fluctuations of the atomic positions, however, deteriorate the three-qubit gate fidelity more severely than the two-qubit gate fidelity. This effect, however, could be reduced by increasing the interatomic interaction strengths.
    If one assumes that those three error sources each contribute an infidelity of $10^{-4}$ to the two-qubit CZ gate, similar to the estimate of Ref.~\cite{tsai2025benchmarking}, we obtain a total CZ-gate error of $1.22\times10^{-3}$. The same fluctuations yield a $\overline{\mathrm{CCZ}}$-gate infidelity of $1.74\times10^{-3}$, which is larger than the two-qubit gate error by a factor of $\approx1.43$.

    Experimental demonstrations of multiqubit Rydberg gates are still rare; however, Evered \emph{et al.}~\cite{evered2023highfidelity} report CZ gate infidelities of $\approx 0.5\%$, while in the same experiment, $\overline{\mathrm{CCZ}}$ gate infidelities are larger by a factor of approximately four. This is significantly more than we expect from the analysis above, suggesting error rates increased by a factor of roughly $1.5$.
    Based on Fig.~\ref{fig:gate_errors}, one may speculate that larger positional fluctuations might contribute to this finding, as the three-qubit gate fidelity is much more sensitive in this regard than the two-qubit gate fidelity.
    Furthermore, the discrepancies between experimentally observed and predicted three-qubit gate infidelities may indicate greater difficulties in calibrating the complex pulses required to implement multiqubit gates. This emphasizes the importance of analytical few-parameter pulses provided, e.g., by \texttt{RydOpt}.

\section{Multiqubit-controlled gates for measurement-free QEC}\label{sec:mf_qec}

    In this section, we investigate the benefit of multiqubit gates for fault-tolerant measurement-free quantum error correction schemes.
    We begin by summarizing how these recently developed protocols work, thereby explaining how three-qubit CCZ gates occur in that context. We then analyze the required CCZ gates in practical scenarios and perform circuit-level noise simulations to estimate the break-even error rates of measurement-free QEC protocols for realistic gate noise.

    \begin{figure*}[t]
        \centering
        \includegraphics[width=\linewidth]{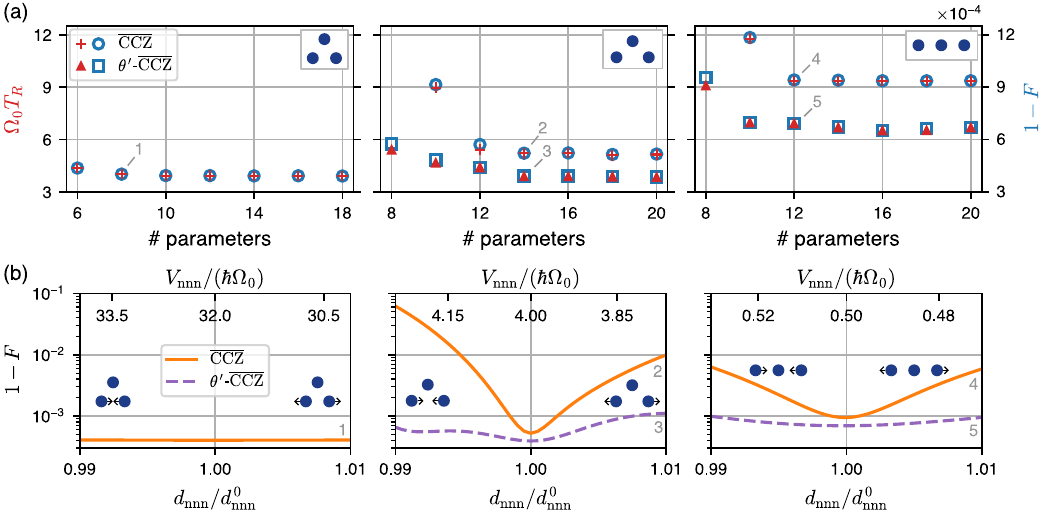}
        \caption{\textbf{$\bm{\overline{\mathrm{CCZ}}}$ gate at finite interaction strengths and asymmetric atomic geometries.} (a) Minimal Rydberg times achievable for $\overline{\mathrm{CCZ}}$ and $\theta'$-$\overline{\mathrm{CCZ}}$ gates on three atoms arranged in an equilateral triangle with $V_{\mathrm{nn}}/(\hbar \Omega_0) = 32$, a right triangle ($V_{\mathrm{nn}}/(\hbar \Omega_0) = 32$, $V_{\mathrm{nnn}}/(\hbar \Omega_0) = 4$), and on a line ($V_{\mathrm{nn}}/(\hbar \Omega_0) = 32$, $V_{\mathrm{nnn}}/(\hbar \Omega_0) = 0.5$). We use the antisymmetric pulse ansatz~\eqref{eq:ansatz_sin} and perform $8\times 10^4$ optimizations for each data point, choosing the best run.
        (b) Sensitivity of the gates indicated above to variations in the interatomic distances, i.e., variations in the interaction strengths. The left plot considers an 8-parameter pulse calibrated for $V_{\mathrm{nn}}/(\hbar \Omega_0) = V_{\mathrm{nnn}}/(\hbar \Omega_0) = 32$. The center panel investigates two 14-parameter pulses calibrated for atoms arranged in a right triangle. The right panel considers two 12-parameter pulses calibrated for atoms arranged on a line. In all panels in this figure, the decay strength is set to $\gamma/\Omega_0=10^{-4}$.}
        \label{fig:CCZ_finiteV_combined}
    \end{figure*}

\subsection{Fault-tolerant measurement-free QEC}

    Traditional fault-tolerant quantum error correction protocols require midcircuit measurements of ancilla qubits to read out the stabilizers and, potentially, also flag qubits. A decoder then processes the extracted classical syndrome information to apply real-time feedback operations to the data qubits.
    Midcircuit measurements, however, are technically challenging in many hardware platforms, such as those based on neutral atoms. In atomic quantum computing platforms, qubits are measured by performing fluorescence imaging.
    In the latest large-scale neutral-atom quantum computing experiments, this process still takes much longer than typical gate times~\cite{bluvstein2025architectural}. If performed naively, fast measurements eject atoms from their traps, and scattered photons might accidentally excite other atoms. However, these challenges are currently being addressed by incorporating experimental developments such as cavity-assisted readout, shelving of atomic populations, and other techniques into state-of-the-art experiments~\cite{deist2022midcircuit, hu2025site, falconi2025microsecond, graham2023midcircuit, lis2023midcircuit, norcia2023midcircuit, huie2023repetitive, chiu2025continuous}.
    Nevertheless, the aforementioned difficulties have led to renewed interest in early proposals~\cite{boykin2010algorithms, pazsilva2010faultolerant, crow2016improved} and motivated further research in the direction of measurement-free fault-tolerant QEC~\cite{heussen2024measurementfree, perlin2023fault, veroni2024optimized, brechtelsbauer2025measurementfree} and also measurement-free universal quantum computing protocols~\cite{butt2024measurement, veroni2025universal} in recent years.

    All these protocols have in common that stabilizer information is mapped onto ancilla qubits and feedback operations are then applied \emph{coherently} to the data qubits. Entropy is removed from the system by resetting the ancilla qubits afterwards or replacing them with fresh ones.
    The coherent feedback operations implement the decoding of classical syndrome information encoded in the ancilla qubits and the resultant correction of errors. This procedure requires multiqubit gates. The protocols from Ref.~\cite{heussen2024measurementfree} that realize fault-tolerant measurement-free QEC on the seven-qubit Steane code employ $\mathrm{C}_n\mathrm{Z}$ gates with $n=2,3$, whereas the schemes in Ref.~\cite{veroni2024optimized} require only $\mathrm{C}_{2}\mathrm{Z}$ gates.
    The protocols from both of these papers have in common that a subset of qubits, on which the multiqubit gates act, is reset afterwards. This can be useful if the spatial arrangement of atoms is not symmetric under the exchange of any two atoms. We discuss this in more detail in the following subsection.

\subsection{CCZ gate for finite interactions}

    In experiments, atoms are usually not arranged at distances close enough to assume an infinite Rydberg blockade, and they might not even be arranged symmetrically.
    Here we discuss the implications of such circumstances, and we will see that the option to reset a subset of qubits after the application of a native $\mathrm{C}_n\mathrm{Z}$ gate can be valuable.

    First, we examine the case of symmetrically arranged atoms in the strong but finite blockade regime.
    As discussed previously, a realistic value for the nearest-neighbour interaction strength is $V_{\mathrm{nn}}/(\hbar \Omega_0) = 32$, which we therefore employ in the following. We optimize $\overline{\mathrm{CCZ}}$ gate pulses in the presence of Rydberg-state decay to determine the shortest possible Rydberg times $\Omega_0T_R$, using the antisymmetric ansatz~\eqref{eq:ansatz_sin}. We find Rydberg times that are very similar to those in the infinite-blockade case, as shown in the left panel of Fig.~\ref{fig:CCZ_finiteV_combined}(a). The eight-parameter pulse marked with the number \textsf{1} is the pulse that we have used in the gate-error analysis of Sec.~\ref{sec:errors_cz_ccz}. Here we employ a Rydberg decay strength $\gamma/\Omega_0=10^{-4}$, which slightly underestimates the realistic decay used earlier, but allows for a convenient extrapolation of gate infidelities given a specific value of $\gamma/\Omega_0$.
    To visualize the robustness of the marked $\overline{\mathrm{CCZ}}$ gate with respect to small variations in interatomic distances, we consider two of the atoms being slightly pushed away from each other or pushed towards each other. The left panel in Fig.~\ref{fig:CCZ_finiteV_combined}(b) indicates that variations of the order of 1\% affect the gate infidelity by an amount that is smaller than $10^{-4}$.
    
\begin{figure}[t]
        \centering
        \includegraphics[width=\linewidth]{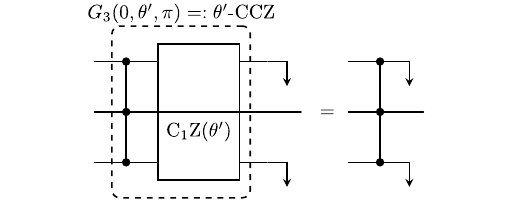}
        \caption{\textbf{CCZ gate followed by reset of two qubits.} The operation can be realized by a global multiatom Rydberg gate \mbox{$G_3(0, \theta', \pi)$} $=: \theta'$-$\mathrm{CCZ}$, where the precise value of the gate parameter $\theta'$ does not matter. In Appendix~\ref{app:more_multiqubit_gates}, Fig.~\ref{fig:circuit_CCZ_fast_eps}, we show that the faster Rydberg gate \mbox{$G_3(\pi, \theta', \pi)$} $=: \theta'$-$\overline{\mathrm{CCZ}}$ can also implement a CCZ gate followed by a reset of two qubits.}
        \label{fig:circuit_ccz_reset}
    \end{figure}

    Next, we investigate CCZ gates applied to three qubits arranged in an isosceles triangle. If two qubits are reset after the gate, which is common in MF QEC protocols, one gains the freedom to implement the CCZ gate up to a well-defined miscalibration that we sketch in Fig.~\ref{fig:circuit_ccz_reset}. If the qubits to be discarded correspond to the atoms labeled 1 and 3, the desired gate can be realized by implementing a global Rydberg gate $G_3(0, \theta', \pi) =: \theta'$-$\mathrm{CCZ}$. An arbitrary value of $\theta'$ does not matter, because the two qubits affected by it will be reset.
    Figure~\ref{fig:circuit_CCZ_fast_eps} in Appendix~\ref{app:more_multiqubit_gates} shows that the implementation of a faster $\overline{\mathrm{CCZ}}$ gate can also be simplified if two qubits are discarded after the gate, resulting in $G_3(\pi,\theta',\pi) =: \theta'$-$\overline{\mathrm{CCZ}}$.
    Note that this freedom in gate design only matters if the atoms are arranged in a configuration that is not totally symmetric. Otherwise, the parameter $\theta'$ cannot be chosen independently from $\theta$.

    We now look at two concrete examples, namely $\overline{\mathrm{CCZ}}$ gates and $\theta'$-$\overline{\mathrm{CCZ}}$ gates applied to three atoms arranged in a right triangle and three atoms arranged on a line. These geometries are likely to occur if atoms are placed on a square lattice, and individual atoms can be addressed with a Rydberg laser~\cite{radnaev2025universal}.
    We keep the interaction strength between nearest neighbours at $V_{\mathrm{nn}}/(\hbar \Omega_0) = 32$. The right triangle then corresponds to a next-nearest neighbour interaction strength of $V_{\mathrm{nnn}}/(\hbar \Omega_0) = 4$, and for the line we get $V_{\mathrm{nnn}}/(\hbar \Omega_0) = 0.5$.
    In both of these cases, the realization of the gate is more difficult than in the infinite blockade and the symmetric finite blockade cases because more constraints have to be fulfilled and more states are involved in the dynamics. We therefore expect the number of pulse parameters required to implement the gate to be higher than in the cases considered so far. This turns out to be true, as shown in the center and right panels of Fig.~\ref{fig:CCZ_finiteV_combined}(a).
    The plots show the Rydberg times $\Omega_0 T_R$ and the infidelities achievable in the presence of Rydberg decay for varying numbers of pulse parameters. We see that the pulses exhibit larger Rydberg times than in the symmetric case analyzed in the left panel.
    Moreover, we observe that the $\theta'$-$\overline{\mathrm{CCZ}}$ gate can be implemented with fewer pulse parameters and smaller Rydberg times $\Omega_0 T_R$ than the traditional $\overline{\mathrm{CCZ}}$ gate.
    A comparison with Rydberg times of time-optimal CZ gates reveals that the $\theta'$-$\overline{\mathrm{CCZ}}$ gates discussed here exhibit values of $\Omega_0 T_R$ that are at most $\approx \! 2.5$ times larger. These gates may thus be very useful for measurement-free QEC applied to atoms arranged in a lattice.
    The second and third plots of Fig.~\ref{fig:CCZ_finiteV_combined}(b) show the robustness of the gates indicated in panel (a) with respect to variations in interatomic distances. The $\overline{\mathrm{CCZ}}$ gates are much more sensitive in this regard than the gate in the symmetric configuration with a strong blockade. Furthermore, one observes that the $\theta'$-$\overline{\mathrm{CCZ}}$ gates are more robust than the full $\overline{\mathrm{CCZ}}$ gates.
    Variations in interatomic distances may be caused by variations of the exact tweezer positions or by forces between atoms excited to Rydberg states. In the strong blockade regime, joint Rydberg excitations are highly suppressed. However, in the linear atomic configuration considered here, the two outermost atoms do not block each other. We therefore expect distance variations to be particularly present in this setup. For three atoms on a line, a large sensitivity of a gate to variations in the interaction strength may thus result in poor gate performance~\cite{kazemi2025multiqubit}.

\subsection{Performance of measurement-free fault-tolerant QEC with native multiqubit gates}

    \begin{figure}[t]
        \centering
        \includegraphics[width=\linewidth]{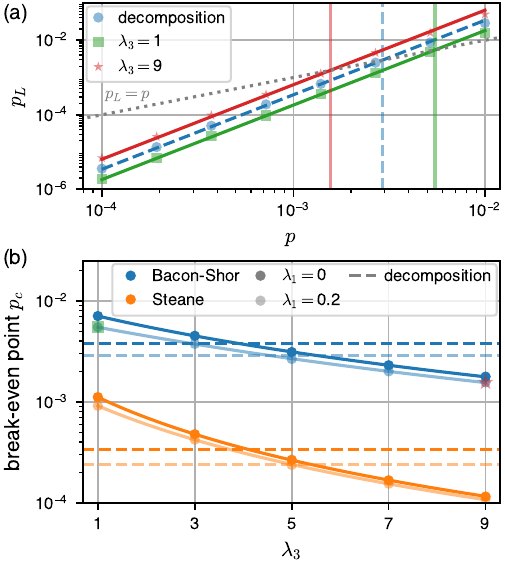}
        \caption{\textbf{Measurement-free QEC performance.} Using protocols from Ref.~\cite{veroni2024optimized} for the Bacon-Shor code and the Steane code, we analyze the logical error rate for various single-, two- and three-qubit gate errors. (a) Error rate $p_L$ of a logical qubit encoded in the Bacon-Shor code as a function of the two-qubit gate error rate $p$ for $p_1/p=0.2$, and $p_3/p=\lambda_3$. Logical error rates are determined by averaging over the input states $\ket{0_L}$ and $\ket{+_L}$. The dashed line corresponds to a decomposition of all three-qubit gates into single- and two-qubit gates. Crossings with the dotted line $p_L=p$, marked also by the vertical lines, define the respective break-even error rates $p_c$. (b) Break-even error rate as a function of the three-qubit gate error factor $\lambda_3$. Lighter shaded colors correspond to $p_1/p=0.2$, darker colors to $p_1/p=0$. Horizontal dashed lines mark the break-even points obtained with decomposed three-qubit gates. Three-qubit error scaling factors $\lambda_3\approx4$ have been demonstrated experimentally~\cite{evered2023highfidelity}, while $\lambda_3\approx1.5$ could be possibly achieved.}
        \label{fig:MF_QEC_results}
    \end{figure}

    \begin{figure*}
        \centering
        \includegraphics[width=\linewidth]{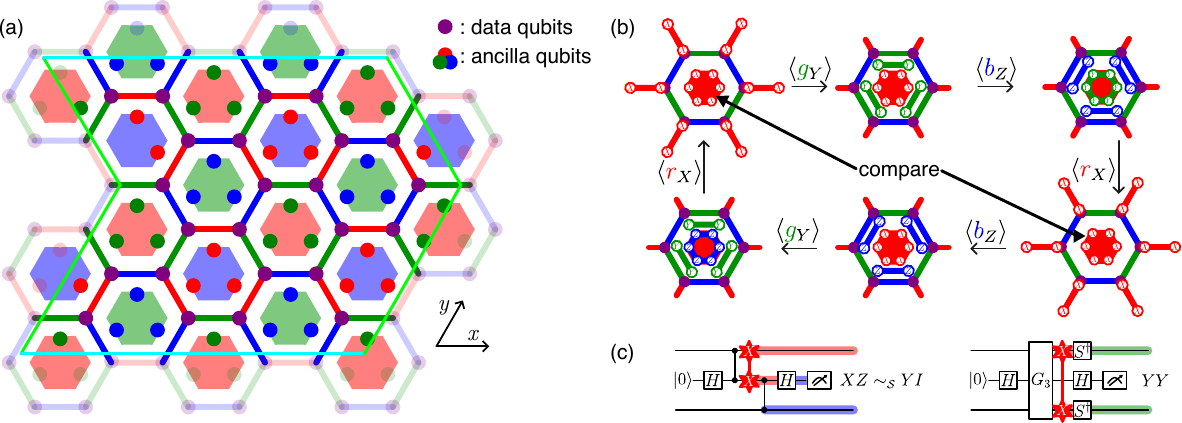}
        \caption{\textbf{The honeycomb Floquet quantum error correction code.} (a) Planar honeycomb Floquet code of size $s=2$. The data qubits are located on the vertices of a hexagonal, tri-colorable tiling. Boundaries cut through edges of the same color along one direction of the torus. In the $x$-direction, we cut blue edges; in the $y$-direction, we cut green edges. In this direction, we also cut a zig-zag boundary as in Ref.~\cite{gidney2022benchmarking}. Red, green, and blue edges indicate a measurement of Pauli $XX$, $YY$, and $ZZ$ on adjacent qubits. In the bulk, every red (green, blue) plaquette admits a $6$-body $X \, (Y, Z)$ stabilizer that commutes with all adjacent measurements. (b) The rewinding measurement schedule periodically measures the sequence $[\langle \redr_{X} \rangle, \langle \greeng_{Y} \rangle, \langle \blueb_{Z} \rangle, \langle \redr_{X} \rangle, \langle \blueb_{Z} \rangle, \langle \greeng_{Y} \rangle]$. We show a red bulk plaquette and for each subround a generating set of stabilizers supported on that plaquette. Note that the $X$-plaquette stabilizer is measured twice---the eigenvalue can always be reconstructed using subsequent $YY$- and $ZZ$-measurements. Once initialized, these outcomes will always be the same due to commutation if there are no faults. This allows for the detection of errors by comparing measurement outcomes of plaquettes to those of previous (sub)rounds. (c) In an implementation with two-qubit gates (left), a correlated $XX$-error after the first entangling gate propagates to an $XZ$-error on the data qubits. This error is stabilizer equivalent to a single-qubit $Y$-error. In a three-qubit gate implementation (right), a correlated $XX$-error can occur directly on the data qubits, leading to an effectively halved circuit-level distance, as discussed in Sec.~\ref{sec:floquet_performance}.}
        \label{fig:floquet_codes}
    \end{figure*}

    We now investigate the performance of measurement-free fault-tolerant QEC protocols from Ref.~\cite{veroni2024optimized} for various single-, two-, and three-qubit gate error rates.
    We consider two exemplary distance-three codes, namely the seven-qubit Steane code~\cite{steane1996error} and the 9-qubit Bacon-Shor code~\cite{bacon2006operator, aliferis2007subsystem}.
    As a simple but meaningful noise model, we employ multiqubit depolarizing noise for every circuit element. The $n$-qubit depolarizing noise channel with error rate $p_n$, acting on a density matrix $\rho$, reads
    \begin{equation}
        \mathcal{N}_n(\rho) = (1-p_n) \rho + \frac{p_n}{4^n-1} \sum_{P} P \rho P , \label{eq:n_qub_depol}
    \end{equation}
    with $P\in\{I,X,Y,Z\}^{\otimes n} \setminus I^{\otimes n}$. We consider the CZ gate error rate as a reference noise rate denoted by $p$, while the single- and three-qubit gate error rates are scaled by the factors $\lambda_1=p_1/p$ and $\lambda_3=p_3/p$, respectively.
    We model the initialization of a qubit in $\ket{0}$ to be followed by a bit flip with probability $p_1$.
    Consecutive single-qubit gates on the same qubit are combined into one individual single-qubit gate, followed by the error channel.
    To assess the performance of the measurement-free QEC protocols for varying physical error rates, we perform circuit-level noise simulations using \texttt{qiskit}~\cite{qiskit}. We determine the logical error rate $p_L$ as a function of the physical error rate $p$. For the calculation of $p_L$, we average over the logical states $\ket{0}_L$ and $\ket{+}_L$ as inputs to the protocol.
    This allows us to identify a break-even error rate $p_L(p_c) = p_c$, which provides an estimate of the gate error rates required for the QEC protocol to be useful.

    \begin{figure*}[t]
        \centering
        \includegraphics[width=\linewidth]{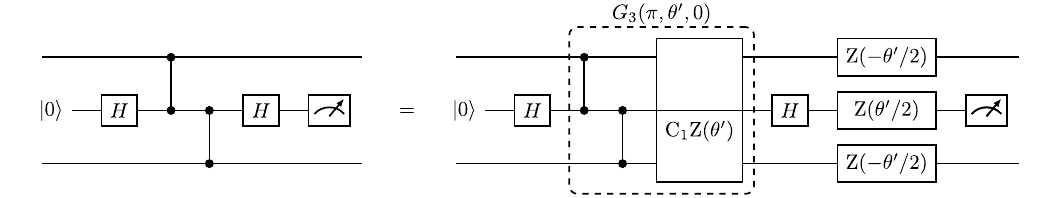}
        \caption{\textbf{Circuits for two-body stabilizer readout.} Both circuits realize the same quantum instrument, described by the Kraus operators $K_0 = \frac{1}{2}(\mathds{1} + Z_1 Z_3)$ and $K_1 = \frac{1}{2}(\mathds{1} - Z_1 Z_3)$ for a measurement of the ancilla qubit in $\ket{0}$ and $\ket{1}$, respectively. In particular, one can use the Rydberg gate $G_3(\pi,\pi,0)$ that is completely symmetric and can thus be implemented in the perfect blockade regime. The gate $\mathrm{Z}(\theta'/2)$ on the ancilla does not affect the measurement and, therefore, can be omitted.}
        \label{fig:circuit_stabilizer_readout}
    \end{figure*}

    Figure~\ref{fig:MF_QEC_results}(a) shows an exemplary scaling of the logical error rate $p_L$ as a function of the physical error rate $p$ for the Bacon-Shor code. The dashed curve exhibits the performance of the protocol if all three-qubit gates are decomposed into single- and two-qubit gates. We decompose three-qubit gates that are followed by the reset of two qubits into four two-qubit and five single-qubit gates, analogous to Ref.~\cite{heussen2024measurementfree}. Two solid curves for different values of $\lambda_3$ show how the break-even error rate $p_c$ decreases for larger three-qubit gate error rates. Figure~\ref{fig:MF_QEC_results}(b) demonstrates this in more detail. Break-even error rates $p_c$ are plotted as a function of $\lambda_3$ for the Bacon-Shor code as well as for the Steane code, for two different single-qubit gate error strengths characterized by $\lambda_1$. Fundamentally, a scaling factor $\lambda_3 \approx 1.5$ should be possible to achieve for three-qubit $\overline{\mathrm{CCZ}}$ gates, as discussed in Sec.~\ref{sec:multiqubit_gates}. In practice, experiments have demonstrated three-qubit gate fidelities that are a factor of $\lambda_3 \approx 4$ worse than the CZ gate fidelities in the same experiment~\cite{evered2023highfidelity}.
    We see that the break-even points for the Bacon-Shor code lie roughly between $2\times 10^{-3}$ and $10^{-2}$. Two-qubit gate error rates in this range have been demonstrated experimentally~\cite{bluvstein2024logical, reichardt2025faulttolerant, bedalov2024faulttolerant, rodriguez2025experimental, bluvstein2025architectural, rines2025demonstration, tsai2025benchmarking}. The Steane code exhibits break-even error rates approximately one order of magnitude lower, which would require improved physical error rates compared to what is available today. These findings align well with the results of Ref.~\cite{veroni2024optimized}, which identified the Bacon-Shor code to exhibit the most promising break-even error rate for near-term hardware.
    The dashed lines in panel (b) correspond to the values of $p_c$ obtained if all three-qubit gates are decomposed into single- and two-qubit gates. This allows us to infer how good the three-qubit gate fidelity must be to outperform the decomposition. We learn that the three-qubit gates may perform approximately four to five times worse than the CZ gates, depending on the single-qubit gate error strength. This can be understood from the fact that a CCZ gate followed by the reset of two qubits has to be decomposed into four CZ gates and five single-qubit gates.
    All in all, we learn that the fidelity requirements for three-qubit gates are not as ambitious as those for the two-qubit gates. For example, CZ gates with $0.3\%$ error rate and CCZ gates with $\approx \! 1.2\%$ error rate could enable break-even demonstrations of measurement-free QEC, which should be feasible with current hardware.

\section{Global stabilizer readout for Floquet codes}\label{sec:floquet}

    This section analyzes the application of global multiqubit gates in the context of Floquet QEC codes. We first briefly introduce Floquet codes and then describe how global three-qubit gates can be used for stabilizer readout in these codes. Finally, we simulate the performance of Floquet QEC cycles using two-qubit gates and three-qubit gates.

\subsection{Floquet quantum error correction}

    Floquet codes, introduced with the \emph{honeycomb code} in Refs.~\cite{hastings2021dynamically, gidney2021faulttolerant}, are dynamical error correcting codes based on periodic low-weight Pauli operator measurements. 
    The honeycomb code is defined on a hexagonal, three-colorable tiling of a torus---we show the planar honeycomb code~\cite{haah2022boundaries, gidney2022benchmarking} in Fig.~\ref{fig:floquet_codes}(a). Qubits are put on vertices, and the colors of the edges define the kind of measurement: Pauli $XX$, $YY$, and $ZZ$ operators are measured on qubits adjacent to red, green, and blue edges, respectively. Equipped with a periodic measurement sequence, as sketched in Fig.~\ref{fig:floquet_codes}(b), the state cycles through different instantaneous stabilizer groups (ISGs). Elements thereof can be periodically reconstructed from the two-body measurements, enabling the detection of errors. The dynamics and error correction properties of such codes can be understood in the detector error model formalism or using graphical calculus~\cite{magdalena2025XYZ, derks2024designing}. We use the construction of Ref.~\cite{gidney2022benchmarking}, which, for a size $s$, results in a family of planar codes with parameters $[[n,k,d]] = [[12s,1,2s-1]]$. For a detailed technical exposition of Floquet codes, we refer to Appendix~\ref{app:floquet}.
    
    A primary motivation for using Floquet codes is that they only require low-weight Pauli measurements. On the one hand, this can reduce the experimental requirements for qubit connectivity when using a planar Floquet code that only requires nearest-neighbor interactions in a 2D plane. On the other hand, the low weight of the measurements eases a fault-tolerant (FT) implementation:
    A two-qubit Pauli measurement realized by a circuit based on two-qubit gates is shown on the left-hand side in Fig.~\ref{fig:circuit_stabilizer_readout}. In a circuit-level noise model with $n$-qubit depolarizing channels after $n$-qubit gates, this circuit is always FT: 
    any fault results in at most one error on the data qubits. 
    This is because (i) any gate involves at most one data qubit, so the corresponding noise channels induce at most a weight-$1$ error; and (ii) the just measured operator, e.g.~a Pauli $ZZ$, is part of the stabilizer group. 
    Take, e.g.~a Pauli $XX$ fault on the first entangling $\mathrm{CZ}$-gate in Fig.~\ref{fig:floquet_codes}(c). This fault propagates to an $XZ$ data error. This has weight two, but multiplying with the (now stabilizer) $(-1)^m ZZ$ results in the weight-$1$ error $\propto YI$.
    
    Another envisioned application of Floquet codes is in the context of proposed platforms based on Majorana zero mode qubits~\cite{paetznick2023performance}.
    In such systems, the two-qubit Pauli measurements can be performed directly without an ancillary qubit. 
    With a general error model that allows for arbitrary two-qubit Pauli errors on the qubits, an $XX$-fault might e.g.~occur while measuring a $ZZ$-operator. This leads to the code distance being asymptotically reduced to half of its original value.
    The smaller number of fault locations, however, results in a much higher threshold. Reference~\cite{gidney2021faulttolerant} reports a threshold of $1.5-2\%$ for the direct two-body measurement compared to $0.2-0.3\%$ for an implementation based on two-qubit $\mathrm{CZ}$ gates.

    \begin{figure*}[t]
        \centering
        \includegraphics[width=\linewidth]{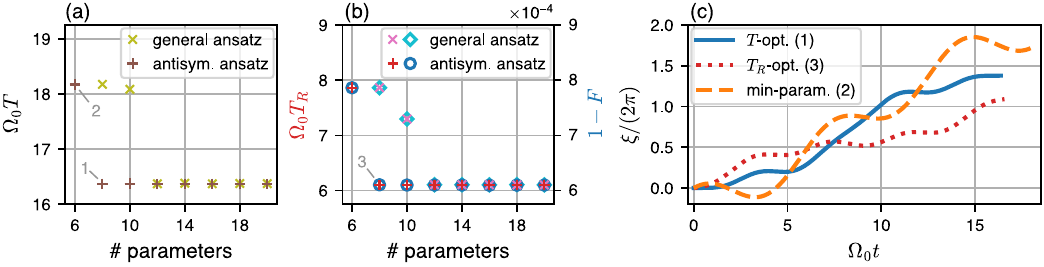}
        \caption{\textbf{Time-optimal and $\bm{T_R}$-optimal CZ-CZ-CZ gates in the perfect Rydberg blockade regime.} For an increasing number of pulse parameters, and thus increasing pulse complexity, we search for CZ-CZ-CZ gates using the general ansatz~\eqref{eq:ansatz_sin_cos} and the antisymmetric ansatz~\eqref{eq:ansatz_sin}. Panel (a) shows the minimal pulse durations $\Omega_0 T$ retrieved among $4\times10^4$ optimization runs for each data point. Every pulse realizes the target gate with infidelity $<\!10^{-7}$ in the \emph{absence} of Rydberg-state decay. The time-optimal pulse requires 8 parameters (\textsf{1}:~$\Omega_0 T = 16.37, \, \Omega_0 T_R = 6.54$). The minimal number of pulse parameters required to realize the gate is 6 (\textsf{2}:~$\Omega_0 T = 18.17, \, \Omega_0 T_R = 7.87$). In panel (b), pulses are optimized in the presence of Rydberg state decay with rate $\gamma/\Omega_0 = 10^{-4}$. In order to achieve the smallest possible gate error, the Rydberg time $T_R$ is minimized. An 8-parameter pulse can reduce $T_R$ compared to the time-optimal gates (\textsf{3}:~$\Omega_0 T = 16.56, \, \Omega_0 T_R = 6.11$). Panel (c) shows the pulse profiles for the gates discussed in panels (a) and (b). All pulse parameters are provided in Appendix~\ref{app:pulse_parameters}, Table~\ref{tab:timeoptimal_CZCZCZ}.}
        \label{fig:CZCZCZ_combined}
    \end{figure*}

    In the following, we explain how the two-body Pauli measurements can be implemented using global three-qubit Rydberg gates. We will see that for depolarizing noise on the multiqubit gates, the circuit-level distance is halved, as for the direct two-qubit Pauli measurements mentioned above. However, the later subsections demonstrate that three-qubit gates may still be beneficial for stabilizer readout in Floquet codes.

\subsection{Stabilizer readout with multiqubit gates}

    The left-hand side of Fig.~\ref{fig:circuit_stabilizer_readout} shows the standard circuit to read out a $ZZ$-stabilizer of a QEC code, such as the honeycomb Floquet code. An ancilla qubit is coupled to two data qubits using two CZ gates, and a final measurement of the ancilla qubit projects the two data qubits into the even or odd subspace of the operator $ZZ$. The circuit thus realizes a quantum instrument with Kraus operators $K_0 = \frac{1}{2}(\mathds{1} + Z Z)$ and $K_1 = \frac{1}{2}(\mathds{1} - Z Z)$ on the data qubits corresponding to a measurement of the ancilla in $\ket{0}$ and $\ket{1}$, respectively.
    The very same quantum instrument can be realized with a slightly different circuit, shown on the right-hand side of Fig.~\ref{fig:circuit_stabilizer_readout}. The effect of an additional $\mathrm{C_{1}Z}(\theta')$ gate on the two data qubits can be cancelled by appropriate single-qubit gates $\mathrm{Z}(-\theta'/2)$ on the data qubits. This means that instead of two addressed two-qubit CZ gates, a single global three-qubit Rydberg gate $G_{3}(\pi,\theta',0) =:$ CZ-CZ-CZ$(\theta')$ may be used to map the stabilizer information onto the ancilla qubit. Local addressing is just required for single-qubit gates.
    The case $\theta' = \pi$ is of particular interest because the gate $G_{3}(\pi,\pi,0)$---equivalent to CZ gates on all pairs of qubits---is completely symmetric on the three atoms and can thus be realized in the perfect blockade regime. We refer to this gate as CZ-CZ-CZ gate. Generalizations of this gate on up to nine atoms have already been demonstrated experimentally~\cite{cao2024multi}.

    In Fig.~\ref{fig:CZCZCZ_combined}, we investigate time-optimal and $T_R$-optimal CZ-CZ-CZ gates in the perfect blockade regime, similarly to how we analyzed $\overline{\mathrm{CCZ}}$ gates in Fig.~\ref{fig:CCZ_combined}.
    Panel (a) shows that a time-optimal CZ-CZ-CZ gate with a pulse duration $\Omega_0 T = 16.4$ can be implemented with an 8-parameter pulse using the antisymmetric pulse ansatz. Panel (b) indicates that eight parameters suffice to realize a $T_R$-optimal gate with Rydberg time $\Omega_0 T_R = 6.1$. In panel (c), we sketch the profiles of the pulses that realize a CZ-CZ-CZ gate in minimal time, minimal Rydberg time, and with the least number of pulse parameters.
    One might wonder how well a CZ-CZ-CZ$(\theta')$ gate can be implemented for atomic geometries that are not symmetric under the exchange of any two atoms. We find that these geometries lead to larger Rydberg times as compared to the infinite blockade regime, as demonstrated in Appendix~\ref{app:more_multiqubit_gates}.
    Two subsequent time-optimal CZ gates exhibit a total duration $\Omega_0 T = 15.2$ and Rydberg time $\Omega_0 T_R = 5.9$, which is only slightly shorter than for the global CZ-CZ-CZ gate. However, if every CZ gate requires a rearrangement of atoms, the global gate could halve the number of shuttle operations.
    Note, however, that a global three-qubit gate may introduce correlated errors on the participating qubits, which affects the fault tolerance of protocols making use of this gate. The following subsection discusses CZ-CZ-CZ gates in the context of honeycomb Floquet codes.
    Furthermore, we note that it is not possible to read out a four-body stabilizer using two consecutive CZ-CZ-CZ gates that involve a single ancilla and four data qubits. The protocols for fault-tolerant stabilizer readout in the unrotated surface code with three-qubit gates discussed in Refs.~\cite{old2025faulttolerant, pecorari2025lowdepth} thus require genuinely asymmetric gates $G_{3}(\pi,0,0)$.

    \begin{figure*}
        \centering
        \includegraphics[width=\linewidth]{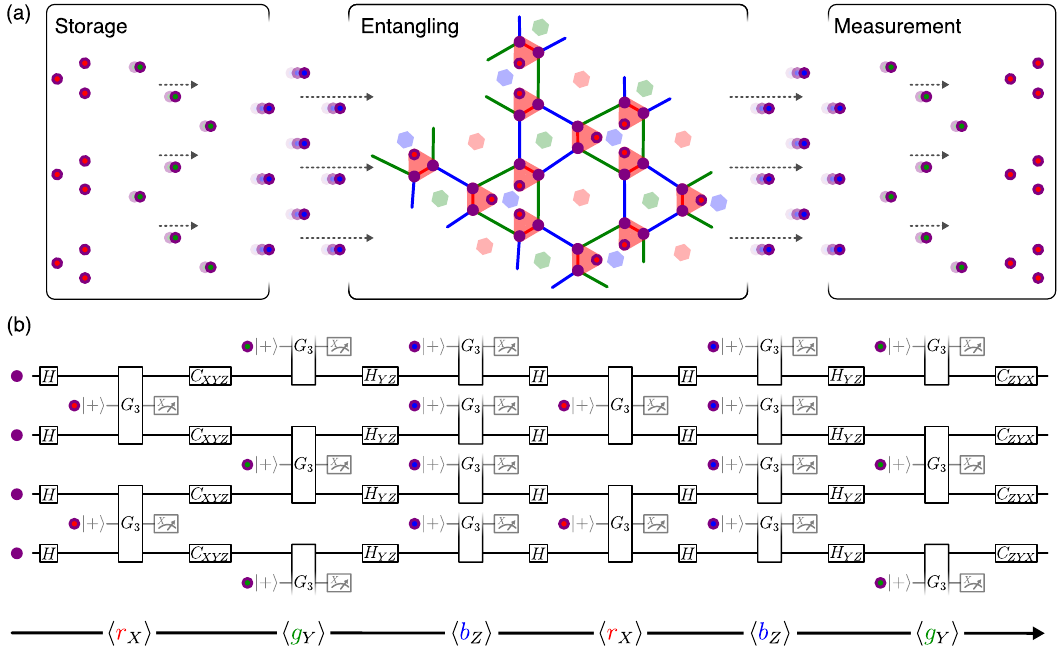}
        \caption{\textbf{Quantum error correction with the honeycomb Floquet code in a neutral-atom platform.} (a) Possible platform architecture consisting of a storage zone, an entangling zone, and a measurement zone. A nearly constant conveyance of ancilla qubits from the storage into the measurement zone passes through the entangling zone to perform the required entangling operations. (b) Circuit for one round of measurements. Note that every data qubit participates in every Floquet measurement subround and thus in every Rydberg laser pulse. This means that data qubits never have to be shuttled out of the entangling zone. Also, the single-qubit gates can be applied globally to all data qubits. We denote $H_{YZ} = \frac{Z+Y}{\sqrt{2}}$, $C_{ZYX} = H_{YZ} H$  and $C_{XYZ} = H H_{YZ}$.}
        \label{fig:platform}
    \end{figure*}

\subsection{Implementation of a Floquet QEC protocol in a neutral atom quantum processor}

    This subsection discusses a possible implementation of planar honeycomb Floquet codes in neutral-atom platforms. 
    We envision an experimental platform with dedicated \emph{storage}, \emph{entangling} and \emph{measurement} zones~\cite{bluvstein2024logical, bluvstein2025architectural}, as depicted in Fig.~\ref{fig:platform}(a).
    We envision to have a reservoir of ancilla atoms in the storage zone, while the atoms that encode the data qubits remain always in the entangling zone.
    For every measurement subround of the Floquet protocol, a group of ancilla atoms of the respective ``color'' is shuttled into the entangling zone, such that the stabilizer values of the corresponding edges can be mapped onto them. These ancillas are then shuttled into the measurement zone for readout, while a new set of ancillas can simultaneously move from the storage into the entangling zone for the next Floquet subround.
    An observation is that every single data qubit participates in every Floquet subround. One can think of the computational dynamics as a nearly constant conveyance of ancilla qubits from the storage into the measurement zone, passing through the entangling zone with a short stop for the entangling operation. We visualize this process in Fig.~\ref{fig:platform}.
    
    An advantage of using global three-qubit gates for the entangling operation is that every data qubit participates in every Rydberg laser pulse, as can be seen in the circuit snippet in Fig.~\ref{fig:platform}(b).
    Using two-qubit gates, however, every CZ gate pulse involves only half of the data qubits. Suppose the set of data atoms not involved in a CZ gate is kept in the entangling zone. These atoms are still temporarily excited to Rydberg states, rendering them susceptible to associated noise processes. Idling data qubits in the entangling zone are thus nearly as prone to errors as data qubits actively involved in CZ gates. To avoid this, one may shuttle idling atoms out of the entangling zone during a CZ gate. However, each Floquet subround consists of two CZ gates acting on disjoint sets of data atoms. This means that in every step, all data atoms must be shuttled out and back into the entangling zone, which incurs a significant shuttling overhead.

    \begin{figure*}
        \centering
        \includegraphics[width=\linewidth]{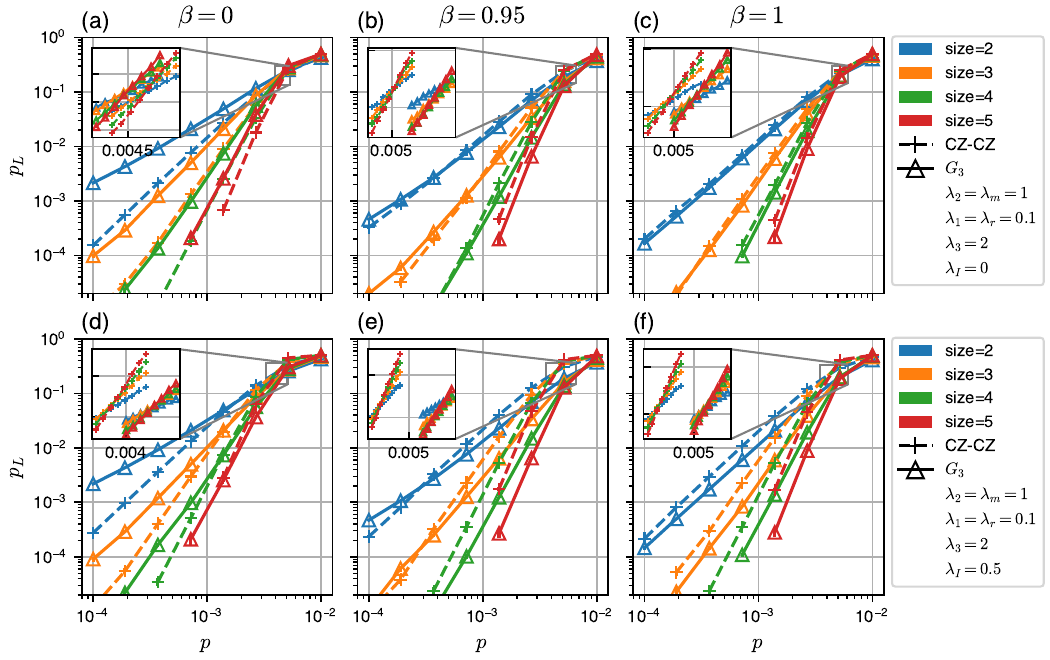}
        \caption{\textbf{Performance of planar honeycomb Floquet codes with two-qubit gates and three-qubit gates.} The two-qubit gate error serves as a reference noise strength $p$. Single-qubit, three-qubit, measurement, and reset error rates are scaled by the factors $(\lambda_1=0.1,\lambda_3=2,\lambda_{\mathrm{m}}=1,\lambda_{\mathrm{R}}=0.1)$, respectively. Different panels show logical error rates $p_L$ for different $Z$-bias strengths $\beta$ (left to right) and for different idling noise strengths during CZ gates $\lambda_I$ (top, bottom). Panels (a)--(c) consider $\lambda_I=0$. In (a), we show results for no $Z$-bias, $\beta=0$, corresponding to uniform depolarizing noise. The thresholds for two- and three-qubit gate implementations are very similar. The multiqubit gate implementation, denoted $G_3$, however, exhibits higher logical error rates and a reduced low-$p$ scaling due to a reduced circuit-level distance. (b) For a strong $Z$-bias, $\beta=0.95$, above $p\approx10^{-3}$, the multiqubit gate implementation has an advantage in logical error rate over the CZ-CZ implementation and a significantly higher threshold. However, due to the reduced circuit-level distance, the curves still flatten out for small values of $p$. (c) In an infinite bias regime, $\beta=1$, we recover the full circuit-level distance for the multiqubit gate implementation, with a higher threshold and a lower logical error rate in the simulated physical error regime.
        In panels (d)--(f), we show results for $\lambda_I=0.5$: Idling data qubits during CZ gates are exposed to errors, which models the effect of a Rydberg gate pulse on single atoms that do not participate in CZ gates but remain in the entangling zone. (d) With this noise model, already at zero $Z$-bias, $\beta = 0$, the three-qubit gate implementation outperforms the two-qubit gate implementation in terms of threshold. The logical error rate is lower for $p > 2\times 10^{-3}$. (e) With increasing $Z$-bias, the advantages get more pronounced, until (f) the FT behaviour is restored for infinite $Z$-bias, $\beta=1$.}
        \label{fig:floquet_z_bias}
    \end{figure*}

\subsection{Performance of Floquet QEC with native multiqubit gates} \label{sec:floquet_performance}

    We now benchmark the memory performance of planar honeycomb Floquet codes using the Clifford circuit simulator \texttt{stim}~\cite{gidney2021stim}. 
    We perform memory experiments that estimate the error-corrected expectation value of a logical operator for a specific number of rounds.
    For a circuit with circuit-level distance $d_{\mathrm{circ}}$, we perform memory experiments consisting of three steps:
    \begin{enumerate}
        \item Initializing all $n$ data qubits in the state $\ket{+}^{\otimes n}$.
        \item Performing $3d_{\mathrm{circ}}$ rounds of measurements, each containing the $6$ subrounds of measurements in the schedule $[\langle \redr_{X} \rangle, \langle \greeng_{Y} \rangle, \langle \blueb_{Z} \rangle, \langle \redr_{X} \rangle, \langle \blueb_{Z} \rangle, \langle \greeng_{Y} \rangle]$.
        \item Measuring all data qubits in the $X$-basis.
    \end{enumerate}
    The circuit-level distance is the number of fault mechanisms required to introduce an undetected logical error. It therefore depends on the actual circuit implementing the measurements and on the noise model.
    For a strictly fault-tolerant circuit, this distance coincides with the distance of the underlying code.
    To perform error correction, we annotate \emph{detectors} and identify which measurements contribute to the \emph{observable}, as detailed in Appendix~\ref{app:floquet_qec}. 
    We decode a detector outcome vector using \texttt{pymatching}~\cite{higgott2025sparseblossom} and record a logical error whenever the predicted observable outcome does not match the measured outcome. 
    We rescale the logical error rate to represent the logical error rate per $d_{\mathrm{circ}}$ rounds with $p_L = 1 - (1-p_{L,3 d_{\mathrm{circ}}})^{1/3}$.
    
    \paragraph{Depolarizing noise:}
    First, we employ a depolarizing circuit-level noise model based on a single parameter~$p$.
    Gates acting on $n$ qubits are followed by $n$-qubit depolarizing channels (cf.~Eq.~\eqref{eq:n_qub_depol}) of strengths $p_n = \lambda_n p$. Similar to Sec.~\ref{sec:mf_qec}, we set $\lambda_2 = 1$, such that the CZ gate error rate serves as a reference. Moreover, we fix $\lambda_1 = 0.1$, since single-qubit gates typically exhibit much smaller error rates than two-qubit gates. We choose $\lambda_3 = 2$, reflecting that the total Rydberg time $\Omega_0 T_R$ of the CZ-CZ-CZ gate is approximately twice as long as for the CZ gate. Resets are followed by a single-qubit depolarizing channel of strength $p_{\mathrm{R}} = 0.1p$ and measurements are preceded by a single-qubit depolarizing channel of strength $p_{\mathrm{m}} = p$.
    
    Figure~\ref{fig:floquet_z_bias}(a) demonstrates the scaling of the logical error rate $p_L$ as a function of the physical error rate $p$ for different code distances. As expected, for the implementation with two-qubit gates, the circuit is distance preserving and $d_{\mathrm{circ}}^{\mathrm{CZ}} = d = 2s-1$. 
    Similar to the noise models inspired by Majorana zero modes~\cite{gidney2022benchmarking, paetznick2023performance}, we find that the circuit-level distance is essentially halved for the three-qubit gate implementation: $d_{\mathrm{circ}}^{G_3} = s$. 
    This is due to arbitrary weight-two Pauli faults that may occur on two data qubits involved in a measurement. The thresholds, however, are very similar for both implementations.

    \paragraph{$Z$-biased noise:}
    The above analysis assumes depolarizing noise on all circuit elements, which is expressive but might be overly pessimistic in some cases. In neutral-atom systems specifically, the errors on two- and multiqubit gates have a special character.
    It has been shown, as discussed above, that the most dominant errors originate from Rydberg state decay~\cite{pagano2022error, tsai2025benchmarking}.
    An evident question is what types of errors are caused by such decay events. In general, a Rydberg decay event can result in three outcomes. The atom may decay to either of the qubit states, $\ket{0}$ or $\ket{1}$, or the atom may end up in neither of the computational states.
    The latter case is called leakage. Leakages may be detected, resulting in localized errors, so-called erasures, which are beneficial for QEC~\cite{wu2022erasure, sahay2023highthreshold, ma2023high}. Another option to address leakage errors is to pump the population in these states into the computational state $\ket{1}$~\cite{cong2022hardware}. Because the two- and multiqubit gate Hamiltonian couples just the state $\ket{1}$ to the Rydberg state, this procedure results solely in $Z$-type errors.
    This is also the case for direct decay from $\ket{r}$ to $\ket{1}$.
    $X$-type errors are only caused by decay events from $\ket{r}$ to $\ket{0}$. It has been argued that for particular atomic species and qubit encodings, these events only make up a small fraction of all errors~\cite{wu2022erasure, cong2022hardware}. Thus, one may encounter the situation in which errors on two- and multiqubit gates are predominantly correlated phase errors.
    Therefore, we investigate the performance of QEC with the honeycomb Floquet code for a $Z$-biased noise model. Concretely, for single-qubit gates, we use a regular depolarizing channel, while on gates acting on $n \geq 2$ qubits we employ an error channel that is biased towards Pauli-$Z$ terms,
    \begin{equation}
        \begin{split}
        & \tilde{\mathcal{N}}^{\beta}_{n}(\rho) = (1-p_n)\rho \\
        & + \frac{p_n}{4^{n}-1} \bigg( \! \left( 1 + 2^{n} \beta \right) \! \sum_{P \in \mathcal{A}_n} \! P \rho P + \left( 1-\beta \right) \! \sum_{P \in \mathcal{B}_n} \! P \rho P \bigg),
        \end{split}
    \end{equation}
    where $\beta$ is a bias parameter and we define $\mathcal{A}_n = \{Z,I\}^{\otimes n}\setminus I^{\otimes n}$, and $\mathcal{B}_n = \{I,X,Y,Z\}^{\otimes n} \setminus \{Z,I\}^{\otimes n}$. For $\beta = 0$, we recover a regular depolarizing error channel, while for $\beta = 1$, we obtain only $Z$-type errors. Note that this convention differs from the one used in Ref.~\cite{brechtelsbauer2025measurementfree}.
    
    Figures~\ref{fig:floquet_z_bias}(b) and (c) show the performance of the discussed Floquet QEC protocols for bias strengths $\beta = 0.95$ and $\beta=1$, respectively. The order of magnitude is inspired by Ref.~\cite{wu2022erasure}, which states that up to $98\%$ of errors might be converted into erasures and, thus, may be converted to $Z$-type errors.
    For finite bias, the reduced distance of the multiqubit gate implementation is clearly visible, indicated by the flattened scaling for small $p$ where $p_L \propto p^{\lfloor(d+3)/4\rfloor}$. 
    A fully fault-tolerant scaling is recovered for infinite $Z$-bias, where the only $2$-qubit faults on the data qubits are $ZZ$-faults corresponding to a stabilizer.
    For the realistic $Z$-bias of $\beta=0.95$, the multiqubit gate implementation can actually outperform the two-qubit gate implementation for relevant physical error strengths of $p > 10^{-3}$. 
    Also, the threshold is higher, $p_{\mathrm{th}}^{G_3} \approx 0.6\% > p_{\mathrm{th}}^{\mathrm{CZ}} \approx 0.5\%$.
    This can be explained by the significant suppression of fault-tolerance-breaking errors in the three-qubit gates. 
    Additionally, there are fewer fault locations. 
    We also observe this influence in the infinite $Z$-bias case. Despite three-qubit gates having twice the noise strength of two-qubit gates, the overall logical error rate is lower, and thresholds are higher.

    \paragraph{Idling errors due to Rydberg excitations:}
    Finally, we analyze the influence of the global Rydberg laser illuminating idling data atoms on QEC performance.
    As explained above, in the two-qubit gate implementation, a single Floquet subround consists of two global CZ gate pulses, each involving half of the data atoms. If the data atoms shall always remain in the entangling zone, the laser thus excites idling atoms to Rydberg states, which are prone to decay. To model this source of error, we apply a $Z$-biased single-qubit depolarizing channel with strength $p_I=0.5p$ to data qubits that are idling during the application of a CZ gate.
    Figure~\ref{fig:floquet_z_bias}(d)--(f) shows the performance of Floquet QEC using this error model with different bias strengths.
    We find that in these cases, the threshold of the three-qubit gate implementation is larger than the threshold of the CZ gate implementation, even in the absence of any noise bias.
    For biased noise, the advantage of multiqubit gates in experimentally relevant noise regimes is even more pronounced than previously discussed.

\section{Conclusion}\label{sec:conclusion}

    In this paper, we have presented the open-source Rydberg gate optimization package \texttt{RydOpt} that generates multiqubit Rydberg gate pulses described by analytic functions. The pulse complexity can be increased systematically by including more pulse parameters.
    We have analyzed three-qubit $\overline{\mathrm{CCZ}}$ and CZ-CZ-CZ gates and found that symmetric pulse ansätze allow us to describe nearly time-optimal and Rydberg time-optimal pulses with only a few parameters. The $\overline{\mathrm{CCZ}}$ gate in particular exhibits a Rydberg time that is only $\approx 30\%$ larger than the Rydberg time of a CZ gate and should therefore be possible to implement with a just slightly lower gate fidelity.
    Multiqubit gates applied to atoms that are not arranged symmetrically are, in general, more complicated to realize than gates on symmetrically arranged atoms. The paper discusses that the respective gate pulses are more complex and exhibit larger Rydberg times. Moreover, we find that these gates are more sensitive to variations in interatomic distances.
    
    We describe how multiqubit gates are highly beneficial for measurement-free fault-tolerant QEC protocols. Even if CCZ gates exhibit error rates approximately four to five times larger than the two-qubit CZ gate error rate, they are more useful than decompositions of multiqubit gates into single- and two-qubit gates.
    For Floquet QEC, we have observed that the use of multiqubit gates may enable the implementation of convenient shuttling schedules, in which ancilla atoms are moved in a conveyor-belt fashion through the entangling zone containing all data atoms. The use of three-qubit gates reduces the circuit-level distance due to fault-tolerance-breaking errors; however, for experimentally motivated biased noise on two- and multiqubit gates, one may see an advantage in logical qubit performance for realistic physical error rates.
    Ultimately, physical error rates cannot be made infinitely small; therefore, the asymptotic scaling of the logical error rate is of rather theoretical interest only. Arbitrarily low logical error rates will be achieved at moderate physical error rates by scaling to large code distances. A better performance of the three-qubit gate implementation for relevant physical error rates may thus be more important than reaching the maximal possible circuit-level distance.

    One of the motivations for the development of measurement-free QEC protocols is that midcircuit measurements are still slow compared to quantum gate operations. In this context, it should be noted that atom shuttling is also considerably slower than gates. However, both the measurement-free QEC and the Floquet QEC protocols discussed in this paper involve significant shuttling moves, either to establish the required qubit connectivity or to move ancilla atoms into a dedicated measurement zone. The margin of benefit of these protocols will therefore depend on details such as shuttling times. Many applications, also beyond those discussed in this paper, would thus benefit from future research aimed at reducing the time overhead associated with atom shuttling.

    In experimental realizations of multiqubit gates, the most dominant errors are likely to result from Rydberg decay, motional dephasing, and, probably, pulse miscalibrations. To achieve the best possible gate performance, one may therefore want to optimize a figure of merit that is susceptible to the Rydberg time $T_R$ and, potentially, to the gate's sensitivity with respect to positional fluctuations. The pulse complexity can be restricted by choosing ansätze with a small number of parameters.
    
    An open question for future research is the feasibility of genuinely non-symmetric multiqubit gates. Recent work~\cite{pecorari2025lowdepth} has demonstrated how $G_3(\pi,0,0)$ gates for stabilizer readout can be implemented with dual species arrays. An interesting question is whether this can be done robustly in single-species atomic arrays, either by spatial blockade engineering or locally addressed phases.
    Another open question is whether multiqubit gates on four or five atoms can be designed such that errors are restricted in a way that makes the gates feasible for fault-tolerant stabilizer readout~\cite{pecorari2025lowdepth, old2025faulttolerant}, or, more generally, for fault-tolerant circuit design.
    Finally, a more explicit inclusion and handling of leakage and loss errors would be a valuable extension of the presented results. An active research direction is, for example, how to advance measurement-free QEC schemes to handle these important types of errors.

    \section*{Data and Code Availability}
    The data shown in this paper is available on Zenodo at~\url{https://doi.org/10.5281/zenodo.17750048}.
    The Rydberg gate optimization package \texttt{RydOpt} is available on GitHub~\cite{rydopt}.

\begin{acknowledgments}
    We thank Johannes Zeiher for fruitful discussions.
    All authors acknowledge support by the German Federal Ministry of Research, Technology and Space (BMFTR) through the project MUNIQC-Atoms (Grant No. 13N16070), and by the German Research Foundation (DFG) through the DFG Priority Programme SPP 2514.
    M.M., D.F.L., J.O., and J.H. acknowledge funding by the Munich Quantum Valley (K-8), which is supported by the Bavarian state government with funds from the Hightech Agenda Bayern Plus, and by the DFG under Germany’s Excellence Strategy ‘Cluster of Excellence Matter and Light for Quantum Computing (ML4Q)' EXC 2004/1 -- 390534769.
    K.B., S.W., and H.P.B. acknowledge funding from the BMFTR through the grants QRydDemo and BeRyQC, and from the Horizon Europe program HORIZON-CL4-2021-DIGITAL-EMERGING-01-30 through the project EuRyQa (Grant No. 101070144).
    The authors gratefully acknowledge the computing time provided to them at the NHR Center NHR4CES at RWTH Aachen University (Project No.~p0020074). This is funded by the BMFTR and the state governments participating based on the resolutions of the GWK for national high-performance computing at universities.
\end{acknowledgments}

\clearpage
\appendix
\onecolumngrid

\section{More multiqubit gates} \label{app:more_multiqubit_gates}

\begin{figure*}[t]
        \centering
        \includegraphics[width=\linewidth]{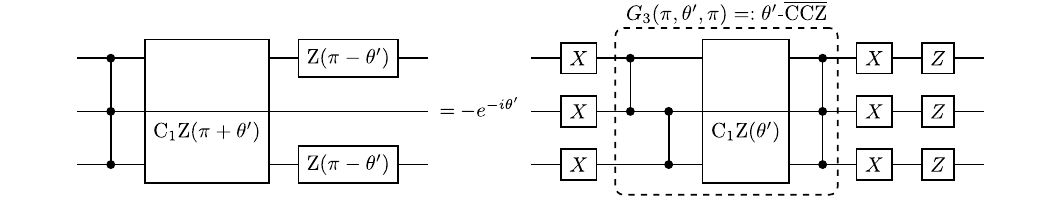}
        \caption{\textbf{Fast implementation of a CCZ gate followed by reset of two qubits.} The gate sequence can be used as a CCZ gate followed by reset of qubits 1 and 3. It can be realized by a fast pulse, analogous to Fig.~\ref{fig:circuit_CCZ_fast}, by employing the gate \mbox{$G_{3}(\pi,\theta',\pi)$} $=: \theta'$-$\overline{\mathrm{CCZ}}$ and global single-qubit rotations, as shown on the right-hand side.}
        \label{fig:circuit_CCZ_fast_eps}
    \end{figure*}

    The circuit in Fig.~\ref{fig:circuit_CCZ_fast_eps} illustrates that a CCZ gate followed by reset of the first and the third qubit can be implemented by a global Rydberg gate $G_3(\pi,\theta',\pi)$ and global single-qubit rotations, where the value of $\theta'$ can be arbitrary. This can be useful if the gate is applied on three qubits that are not arranged symmetrically, as discussed in Sec.~\ref{sec:mf_qec} of the main text.

\begin{figure*}[t]
        \centering
        \includegraphics[width=\linewidth]{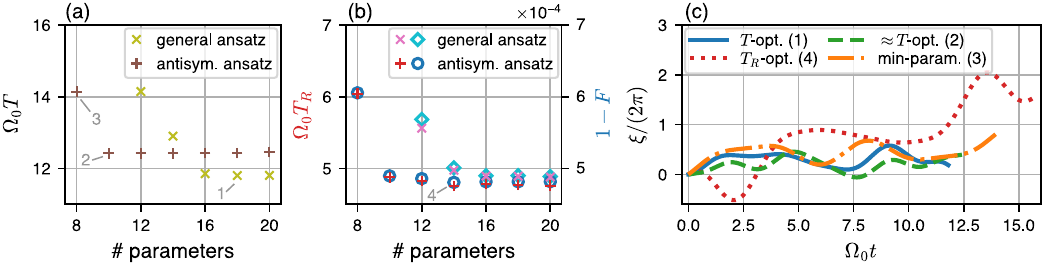}
        \caption{\textbf{Time-optimal and $\bm{T_R}$-optimal $\bm{\overline{\mathrm{CCCZ}}}$ gates in the perfect Rydberg blockade regime.} For an increasing number of pulse parameters, and thus increasing pulse complexity, we search for $\overline{\mathrm{CCCZ}}$ gates using the general ansatz~\eqref{eq:ansatz_sin_cos} and the antisymmetric ansatz~\eqref{eq:ansatz_sin}. Panel (a) shows the minimal pulse durations $\Omega_0 T$ found among $8\times10^4$ optimization runs for each data point. Every such pulse realizes the target gate with infidelity $<10^{-7}$ in the \emph{absence} of Rydberg-state decay. The time-optimal pulse~\cite{evered2023highfidelity} requires 18 parameters (\textsf{1}:~$\Omega_0 T = 11.80, \, \Omega_0 T_R = 5.49$). A nearly time-optimal gate can be realized with a pulse described by 10 parameters (\textsf{2}:~$\Omega_0 T = 12.42, \, \Omega_0 T_R = 4.93$). The minimal number of pulse parameters required to realize the gate is 8 (\textsf{3}:~$\Omega_0 T = 14.14, \, \Omega_0 T_R = 6.41$). In panel (b) pulses are optimized in the presence of Rydberg state decay with rate $\gamma/\Omega_0 = 10^{-4}$. In order to achieve the smallest possible gate error, the Rydberg time $T_R$ is minimized. A 14-parameter pulse can reduce $T_R$ compared to the time-optimal gates (\textsf{4}:~$\Omega_0 T = 15.65, \, \Omega_0 T_R = 4.76$). Panel (c) shows the pulse profiles for the gates discussed in panels (a, b). All pulse parameters are provided in Table~\ref{tab:timeoptimal_CCCZ}.}
        \label{fig:CCCZ_combined}
    \end{figure*}

    Figure~\ref{fig:CCCZ_combined} investigates the four-qubit $\overline{\mathrm{CCCZ}}$ gate in the perfect blockade regime. Similar to the gate $\overline{\mathrm{CCZ}}=G_3(\pi,\pi,\pi)$, we define the gate $\overline{\mathrm{CCCZ}}$ as a unitary consisting of CZ, CCZ, and CCCZ gates acting on all possible subsets of qubits. Panels (a) and (b) show how the minimal gate duration $\Omega_0 T$ and minimal Rydberg time $\Omega_0 T_R$ of the gate depend on the number of pulse parameters.
    Panel (c) plots pulse profiles for gates marked in the previous panels.

\begin{figure}[t]
        \centering
        \includegraphics[width=0.45\linewidth]{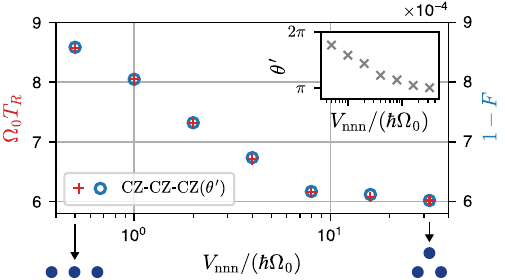}
        \caption{\textbf{CZ-CZ-CZ$\bm{(\theta')}$ gate infidelity and minimal Rydberg times $\bm{\Omega_0 T_R}$ for varying atomic geometries.} We optimize pulse parameters in the presence of Rydberg state decay with rate $\gamma/\Omega_0 = 10^{-4}$, considering the antisymmetric pulse ansatz~\eqref{eq:ansatz_sin} with 12 parameters. The interaction between nearest neighbours is set to $V_{\mathrm{nn}}/(\hbar \Omega_0)=32$, while interaction strengths between next-nearest neighbours are varied from $V_{\mathrm{nnn}}/(\hbar \Omega_0) = 0.5$ (line) to $V_{\mathrm{nnn}}/(\hbar \Omega_0)=32$ (equilateral triangle).}
        \label{fig:CZCZCZ_TR_vs_Vnnn}
    \end{figure}

    Finally, Fig.~\ref{fig:CZCZCZ_TR_vs_Vnnn} analyzes $G_{3}(\pi,\theta',0) =:$ CZ-CZ-CZ$(\theta')$ gates for finite interaction strengths between the three atoms. The plot shows the minimal Rydberg times achievable with 12-parameter pulses for various next-nearest neighbour interaction strengths $V_{\mathrm{nnn}}$, while keeping the nearest neighbour interaction strength constant at $V_{\mathrm{nn}}/(\hbar \Omega_0) = 32$. We vary the atomic geometries from a line to an equilateral triangle. It is visible from the plot that the shortest possible Rydberg time is achieved for the symmetric geometry.
    The inset shows the values of $\theta'$ the respective gates implement. For $V_{\mathrm{nn}}=V_{\mathrm{nnn}}$ it is fixed to be $\pi$. For $V_{\mathrm{nnn}} < V_{\mathrm{nnn}}$, $\theta'$ smoothly deviates from $\pi$.

\section{Details on Floquet codes} \label{app:floquet}

A Floquet code is defined by a periodic sequence of sets of Pauli operators $\mathcal{M} = [\mathcal{M}_i]_{i=0}^{T-1}$. Each $\mathcal{M}_i$ typically consists of single- or two-body Pauli operators and generates a stabilizer group. 
In a Floquet code protocol, the Pauli operators of each sequence are measured one after another.
Typically, we start the Floquet protocol in a product state, e.g. $\ket{+}^{\otimes n}$ which is stabilized by $\mathcal{S}_{-1} = \langle \{X\}_{i=1}^n  \rangle$.
Measuring all operators in $\mathcal{M}_0$ projects the state into the subspace stabilizers by $\{(-1)^{m_i} M_i | M_i \in \mathcal{M}_0\}$.
The new stabilizer group $\mathcal{S}_{0}$ can be efficiently constructed from $\mathcal{S}_{-1}$ and $\mathcal{M}_{0}$~\cite{gottesman1998the}. For each Pauli operator $M_i$ it may anticommute with some $S \in \mathcal{S}_{0}$. 
This set of operators $S \subset \mathcal{S}_0$ can be reduced to a single representative by multiplying all $S_j = S_0 S_j \forall j \in [1,\abs{S}-1]$. 
The only anticommuting element remaining, $S_0$, is then replaced by the just measured operator $(-1)^{m_i} M_i$. 

The measurement sequence therefore defines a sequence of so called \emph{instantaneous stabilizer groups} (ISGs)
\begin{align}
    \mathcal{S}_{-1} \xrightarrow{\mathcal{M}_0} \mathcal{S}_{0} \xrightarrow{\mathcal{M}_1} \mathcal{S}_{1} \xrightarrow{\mathcal{M}_2} \mathcal{S}_{2} \xrightarrow{\mathcal{M}_3} \dots  
\end{align}
In order to preserve the logical information from one ISG to the next, the ISGs have to constitute a \emph{reversible pair of Pauli stabilizer groups}, following Ref.~\cite{aasen2023measurement}. 
One key condition for two ISGs $\mathcal{S}_i$ and $\mathcal{S}_{i+1}$ to constitute such a reversible pair is that they share logical operators.
Formally, there exist a group $\mathcal{L}$ such that $\mathcal{LA} = \mathcal{N}_\mathcal{P}(\mathcal{S}_0)$ and $\mathcal{LB} = \mathcal{N}_\mathcal{P}(\mathcal{S}_1)$.
In practice, only consecutive ISGs share representatives of logical operators. To track the logical operators, the measurement outcomes of edges have to be taken into account. In the following, we explain the establishment of the ISG and the evolution of logical operators in the planar honeycomb Floquet code.

\subsection{Planar honeycomb Floquet code}\label{app:floquet_planar}
The \emph{honeycomb} Floquet code, introduced in Refs.~\cite{hastings2021dynamically, gidney2021faulttolerant} defines the Floquet code on a hexagonal, three-colorable tiling of a torus: qubits are put on vertices, and the colors of the edges define the measurement sequence: Pauli $XX$, $YY$ and $ZZ$ operators are measured on qubits adjacent to red, green and blue edges respectively. We denote by $\langle c_{P} \rangle$ the measurement of Pauli $P$ on all edges of color $c$, then $\mathcal{M} = [\langle \redr_{X} \rangle, \langle \greeng_{Y} \rangle, \langle \blueb_{Z} \rangle]$. 

We now explain how the ISGs of the honeycomb code establish, exemplary for a blue plaquette in the honeycomb Floquet code in Fig.~\ref{fig:floquet_codes}(b).
As any measurement outcome is completely random (the currently measured edges always anticommute with the previously measured edges), we assume without loss of generality that their measurement always yields $+1$.
After the $XX$-measurement of red edges, the corresponding two-body stabilizers are part of the ISG. 
Measuring $YY$ on the green edges, the red edges individually anticommute with the measurements. 
However, their product $X^{\otimes 6}$ commutes with all green $YY$-egdes and is part of the following ISG. 
The subsequent $ZZ$-measurement on (outgoing) blue edges anticommutes with all green edges and the red plaquette. 
Again, their product, now a $Z^{\otimes 6}$ plaquette commutes with the $ZZ$-measurements and is also part of the next ISG. 
Note that after this plaquette has been initialized for the first time, it is always part of the ISG: green $YY$ and red $XX$ edges overlap on two locations and the outgoing blue $Z$ edges trivially commute with the $Z^{\otimes 6}$ plaquette.
Correspondingly, on red and green plaquettes, $X^{\otimes 6}$ and $Y^{\otimes 6}$ operators are always part of any ISG.
The full sequence of ISGs is generated by
\begin{align}
    \langle \langle \redr_X\rangle , \langle {\color{red}{P}}_X\rangle, \langle {\color{mygreen}{P}}_Y\rangle , \langle {\color{blue}{P}}_Z\rangle \rangle \to  
    \langle \langle \greeng_Y\rangle , \langle {\color{red}{P}}_X\rangle, \langle {\color{mygreen}{P}}_Y\rangle , \langle {\color{blue}{P}}_Z\rangle \rangle \to  
    \langle \langle \blueb_Z\rangle , \langle {\color{red}{P}}_X\rangle, \langle {\color{mygreen}{P}}_Y\rangle , \langle {\color{blue}{P}}_Z\rangle \rangle \to \dots  
\end{align}
where we denote by $\langle {\color{red}{P}}_X\rangle$ the set of all 6-body Pauli $X$ operators on red plaquettes.

Focusing on neutral atom platforms, we consider \emph{planar} honeycomb codes~\cite{haah2022boundaries}.
We follow the construction of Ref.~\cite{gidney2022benchmarking}, which requires a rewinding schedule~\cite{dua2023engineering}: $\mathcal{M} = [\langle \redr_{X} \rangle, \langle \greeng_{Y} \rangle, \langle \blueb_{Z} \rangle, \langle \redr_{X} \rangle, \langle \blueb_{Z} \rangle, \langle \greeng_{Y} \rangle]$.
The boundaries are constructed by cutting edges of the same color. In one direction of the torus, they are cut linearly, in a second in a zig-zag fashion. We choose to cut along blue and green edges in the $x$- and $y$-direction, respectively, see Fig.~\ref{fig:floquet_codes}(a).
Cut edges represent single-body measurements. 

\begin{figure}
    \centering
    \includegraphics[width=\linewidth]{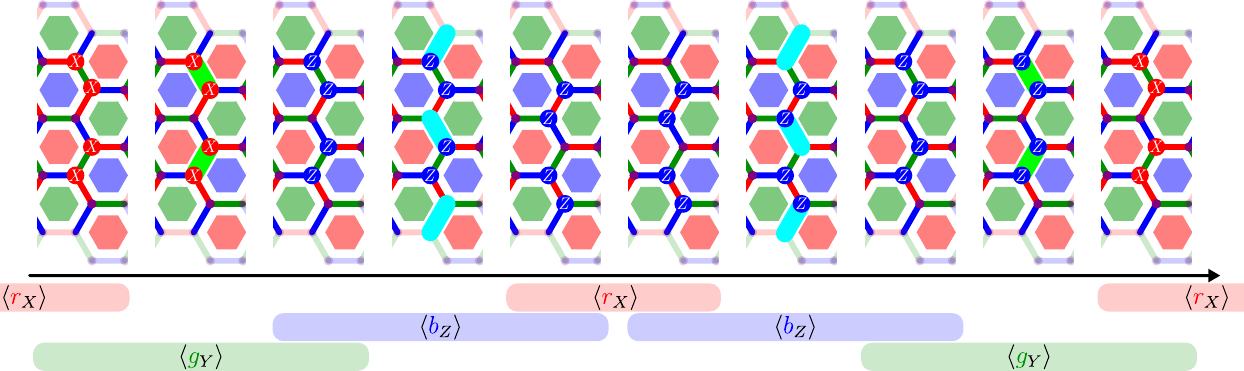}
    \caption{\textbf{Evolution of a vertical logical operator in the planar honeycomb Floquet code.} Time goes from left to right, and we show representatives of logical operators that are shared between subrounds by a shading behind the labels for the measurement. We start with a representative after having measured $\langle \redr_X \rangle$-edges, that is given by Pauli $X$ on green edges along a string in the vertical direction. Because of the even overlap with $\langle \greeng_Y \rangle$-edges, this logical operator is already shared with this subsequent round. The then following $\langle \blueb_Z \rangle$-edges, however, anticommutes with this representative. A shared logical operator is given by multiplying the just measured $\langle \greeng_Y \rangle$-edges, indicated by light green color. This results in a $Z$-string along the green edges. For the following $\langle \redr_X \rangle$-measurement, the blue edges along the string have to be multiplied in the logical operator. At this point, the schedule rewinds and we can follow the same transitions until the next $\langle \redr_X \rangle$-measurement.
}
    \label{fig:logical_evolution}
\end{figure}

In Fig.~\ref{fig:logical_evolution}, we show the evolution of a vertical logical operator. A representative after having measured $\langle \redr_X \rangle$ edges is given by Pauli $X$ on green edges along a string in the vertical direction. Because of the even overlap with $\langle \greeng_Y \rangle$-edges, this logical operator is already shared with the next round. The then following $\langle \blueb_Z \rangle$-edges, however, anticommute with this representative. A shared logical operator is given by multiplying the just measured $\langle \greeng_Y \rangle$-edges, resulting in a $Z$-string along the green edges. For the following $\langle \redr_X \rangle$-measurement, the blue edges along the string have to be multiplied in the logical operator. At this point, the schedule rewinds and we can follow the same transitions until the next $\langle \redr_X \rangle$-measurement.

\subsection{Error correction in planar Floquet codes} \label{app:floquet_qec}

Error correction in general works by looking at measurements or sets of measurements that are deterministic in the absence of noise. 
We refer to such a (set of) measurements as a \emph{detector}~\cite{gidney2021stim, magdalena2025XYZ, derks2024designing}.
For a regular stabilizer code, the detectors are the stabilizer measurements, or parities, i.e. sums, of consecutive stabilizer measurements. 

\begin{figure}
    \centering
    \includegraphics[width=\linewidth]{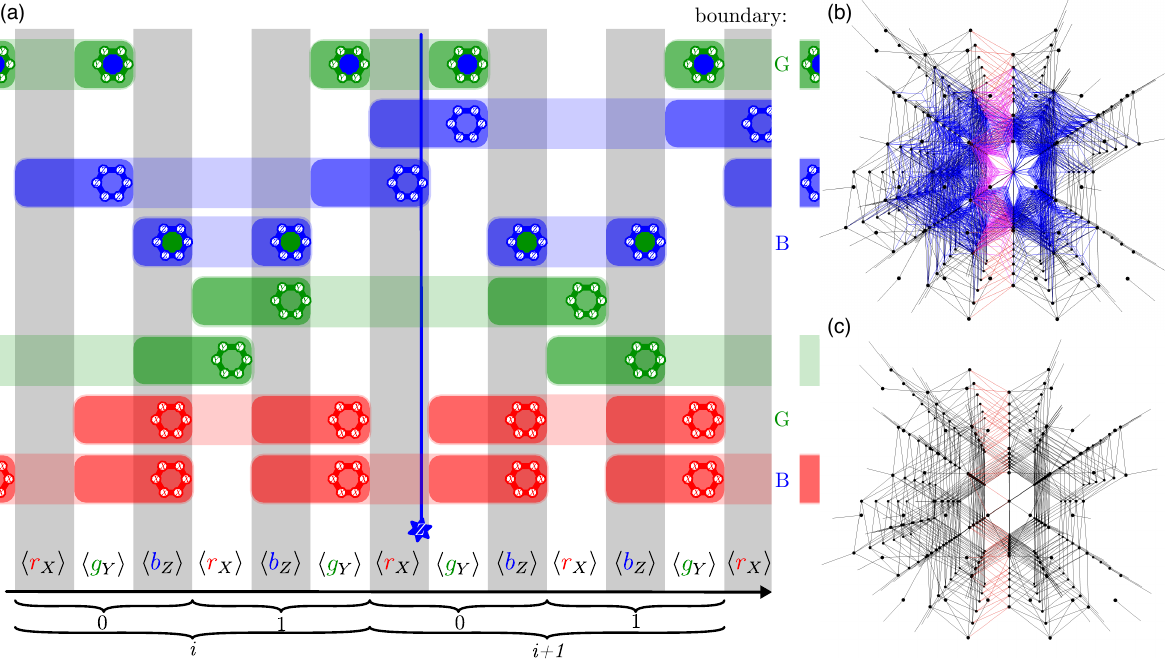}
    \caption{\textbf{Detector error model details.} (a) Detecting regions of the planar honeycomb code. Time goes from left to right and we show rounds $i$ and $i+1$ with their blocks $0$ and $1$. Dark shaded regions indicate steps where measured edges contribute to the drawn plaquette operator. Each detection region starts and ends with a measurement of these operators. Any fault anticommuting with these operators in between subsequent measurement can be detected, indicated by the light shade. As an example, a $Z$-fault after having measured $\langle \redr_X \rangle$ edges is detected by a red $X$-type and two green $Y$-type detectors. (b) Detector error model for a size $s=3$ planar honeycomb Floquet code using a circuit level noise model and three-qubit $G_3$ gates. Detector are represented by circles. Faults correspond to edges. A fault also flipping a logical observable is drawn in red. If a fault flips more than two detectors, the hyperedge is drawn in blue. (c) In the decomposed detector error model, each fault flips at most two detectors, i.e. this is a matchable graph.  }
    \label{fig:detectors_bulk}
\end{figure}

For a Floquet code, typically all two-body Pauli operator measurements are individually random. However, $6$-body $X$ ($Y$,$Z$) Pauli operators on red (green, blue) bulk plaquettes commute with all edges. Once initialized (i.e. the adjacent edges are measured for the first time), they are always part of the ISGs. 
Their eigenvalue can therefore be periodically reconstructed each time the blue and green (blue and red, green and red) edges are measured one after another. 
We visualize these bulk detectors in Fig.~\ref{fig:detectors_bulk}(a). 
By a dark shade, we indicate the rounds where measurements contribute to a plaquette eigenvalue. The light shade spans between two consecutive measurements of plaquette operators. We refer to the total shaded area as a \emph{detecting region}.
A red (i.e. $X$-type) detecting region is sensitive to $Y$- and $Z$-faults occurring in that region, because a $Z$-fault, e.g., anticommutes with the $X$-plaquette and therefore flips its eigenvalue.
Note that every timestep is covered by at least two detecting regions of different colors. 
In practice, a generating set of detectors is necessary, and any basis of detectors is valid. We therefore combine some of the detectors that include $12$ measurements to smaller ones that only require $6$. One example is a detector on green plaquettes consisting of only (blue) $ZZ$-measurement outcomes. This is possible, because in the rewinding schedule, the sequence $\langle \blueb_Z \rangle \to \langle \redr_X \rangle \to \langle \blueb_Z \rangle $ exists. On green plaquettes, the product of $Z$-edges survives the $XX$ measurement. The product around the same green plaquette of immediately following $ZZ$-measurements therefore yields the same outcome compared to two subrounds before.
This kind of detector also exists for blue plaquettes with the corresponding $YY$ measurements in the subsequence $\langle \greeng_Y \rangle \to \langle \redr_X \rangle \to \langle \greeng_Y \rangle $.

At the boundary, qubits undergo single-qubit $Z$- and $Y$-measurements. Here, the order is important: Only if a two-qubit measurement follows two-single-qubit measurements (on the same two qubits), the product of the two single-qubit measurement continues to be part of the stabilizer group. A following single qubit measurement therefore can close a detecting region. 
This is the case for all green and blue boundary plaquettes based on the subrounds $\langle \blueb_Z \rangle \to \langle \redr_X \rangle  \to \langle \blueb_Z \rangle $ and $\langle \greeng_Y \rangle \to \langle \redr_X \rangle \to  \langle \greeng_Y \rangle $ respectively. 
The other detectors on blue and green bulk plaquettes start with red $X$ measurements, which at the boundary get unrecoverably randomized by the subsequent single-qubit $Y$ and $Z$ measurements.
For the red plaquettes, on the green boundary, detectors only exist for detecting regions starting in block 0 ( $\langle \greeng_Y \rangle \to \langle \blueb_Z  \rangle \to \dots $). On the blue boundary, detectors only exist for detecting regions starting in block 1 ( $\langle \blueb_Z \rangle \to \langle \greeng_Y \rangle \to \dots $).
In the corners, no detectors are formed because the corner-qubits are measured in single-qubit $Y$ and $Z$.

In a \emph{memory} experiment, we initialize a specific logical eigenstate and probe whether the logical information survives a given number of error correction cycles. For a CSS stabilizer code, this could e.g. be a single-qubit initialization $\ket{+}^{\otimes n}$ which is a $+1$ eigenstate of the logical $X$-operator. After $n_{\mathrm{rounds}}$ rounds of stabilizer measurements, a single-qubit $X$-measurement allows one to reconstruct the eigenvalue of the logical $X$-operator. The sets of measurements that contribute to the eigenvalue of the logical operator we probe are referred to as observable.
In Floquet codes, the logical operators typically transform from one measurement to the next by multiplication of edge measurement outcomes, see Sec.~\ref{app:floquet_planar}. These measurements therefore also contribute to the logical observable of the memory experiment. 

Finally, given a circuit and a noise model, the action of elementary faults on detector outcomes is specified in a \emph{detector error model}, essentially a collection of matrices $D \in \mathbb{F}_2^{n_{\mathrm{detectors}} \times n_{\mathrm{faults}}}$, $O \in \mathbb{F}_2^{n_{\mathrm{observables}} \times n_{\mathrm{faults}}}$ and a vector  $\vec{p} \in [0,1]^{n_{\mathrm{faults}}}$. 
$D$ is the \emph{detector matrix} encoding which faults flip which detector and can be thought of as the parity check matrix of the circuits.
$O$ is the \emph{observable matrix} that encodes which faults flip the observable(s).  
$\vec{p}$ is the vector of probabilities of the faults.
We show a graphical representation of the detector error model for the circuit-level noise and using the $G_3$-gate in Fig.~\ref{fig:detectors_bulk}(b). There, edges are faults that flip the adjacent detectors, represented by circles. If a fault also flips an observable, it is colored in red. Faults that flip more than two detectors are colored blue.

The simulation then proceeds as follows.
\begin{enumerate}
    \item Sample the noisy circuits. A shot samples a fault-vector $\vec{f}$ according to the probability distribution $\vec{p}$ and calculates the syndrome $\vec{d} = D \vec{f}$ and the observable outcome $\vec{o} = O \vec{f}$.
    \item Decode the syndrome $\vec{d}$. This results in a fault-guess $\vec{f}'$ and an associated observable prediction $\vec{o}\,' = O \vec{f}'$
    \item If $\vec{o}\,' \neq \vec{o}$, count a logical error.
\end{enumerate}

It turns out that decoding the honeycomb Floquet code is equivalent to decoding the surface code~\cite{hastings2021dynamically}.
For a circuit-level noise model, the detector error model still has to be decomposed in order to generate a matchable decoding problem. We show a decomposed detector error model graph in Fig.~\ref{fig:detectors_bulk}(c).
This graph is then suitable for decoding using minimum-weight perfect matching. For all simulations, we use \texttt{pymatching}~\cite{higgott2025sparseblossom}.

\subsection{Other variants of Floquet codes}
There exist CSS versions of honeycomb codes where only $X$- and $Z$- operators are measured~\cite{davydova2023floquet, kesselring2024anyon}.
Planar honeycomb codes were introduced in Refs.~\cite{vuillot2021planar, haah2022boundaries} and benchmarked in Ref.~\cite{gidney2022benchmarking}.
There are numerous other constructions of Floquet codes, \emph{hyperbolic} honeycomb codes, e.g. allow for a higher encoding rate~\cite{higgott2024constructions,fahimniya2025fault}. XYZ ruby codes and variants implement a color code and enable transversal logical gates~\cite{dua2023engineering, davydova2024quantum, magdalena2025XYZ}.

\subsection{Implementing three-qubit gates and $n$-qubit depolarizing errors in \texttt{stim}}
To implement the $G_3$-gates and depolarizing channels acting on $n > 2$ qubits in \texttt{stim}, we place three $\mathrm{CZ}$-gates in the same \texttt{TICK}. To mimic a three-qubit depolarizing channel, we use \texttt{stim}s included correlated error feature. 
We present the basic idea with an example.
Consider a $2$-qubit bit-flip channel with probabilities $p(XI)=p_{XI}, p(IX)=p_{IX}, p(XX)=0$ and $p(II) = 1-p_{XI}-p_{IX}$.
The probability distribution of this channel cannot be reconstructed using \emph{independent} single-qubit channels, because then $p(XX) = p_{IX}p_{XI} + p_{XX} \neq 0$. With the correlated error feature, this $2$-qubit bit-flip channel can be written as
\begin{align}
    \texttt{CORRELATED\_ERROR(p\_XI) X1}\phantom{.} \nonumber \\
    \texttt{ELSE\_CORRELATED\_ERROR(p\_IX/(1-p\_XI)) X2}.
\end{align}
For $p_{IX} = p_{XI} = 0.01$, this is
\begin{align}
    \texttt{CORRELATED\_ERROR(0.01) X1}\phantom{.} \nonumber \\
    \texttt{ELSE\_CORRELATED\_ERROR(0.01010101) X2}. 
\end{align}
In the simulation, the \texttt{ELSE\_CORRELATED\_ERROR(p) P1*P2*\dots} only occurs with probability $p$ if none of the preceding \texttt{(ELSE\_)CORRELATED\_ERROR}s occurred. 
For a non-uniform depolarizing channel, e.g. the $Z$-bias channel from above, on $n$ qubits with probabilities $\vec{p} = \{p_i\}_{i=1}^{4^n-1}$ of (Pauli) errors $\vec{E} = \{P_i\}_{i=1}^{4^n-1}$, we can therefore simulate the channel using subsequent and ordered \texttt{ELSE\_CORRELATED\_ERROR($\tilde{p}_j$)}s with rescaled probabilities,
\begin{align}
    \tilde{p}_i = \frac{p_i}{\prod_{j<i} (1-\tilde{p}_j)} = \frac{p_i}{1 - \sum_{j<i} p_j}.
\end{align}

\section{Pulse parameters} \label{app:pulse_parameters}

    This Appendix provides parameters for the pulses whose profiles are shown throughout this paper.
    Table~\ref{tab:timeoptimal_CZ} contains information on the time-optimal and $T_R$-optimal two-qubit CZ gates.
    Table~\ref{tab:timeoptimal_CCZ} presents the parameters for the time-optimal, $T_R$-optimal, and minimal-parameter pulses implementing three-qubit $\overline{\mathrm{CCZ}}$ gates.
    Table~\ref{tab:timeoptimal_CCCZ} provides data for four-qubit $\overline{\mathrm{CCCZ}}$ gates discussed in Appendix~\ref{app:more_multiqubit_gates}.
    Finally, Table~\ref{tab:timeoptimal_CZCZCZ} presents the pulse parameters for three-qubit CZ-CZ-CZ gates.

    \begin{table*}[t]
    \begin{tabular}{c|r|r} \hline \hline
        & Ansatz~\eqref{eq:ansatz_sin} & Ansatz~\eqref{eq:ansatz_sin} \\
        & 4 parameters & 6 parameters \\ \hline
        $\Omega_0 T$        & 7.61140652  & 7.72506187  \\
        $\Delta_0/\Omega_0$ & 0.07842706  & -0.92491109 \\
        $A_1$               & 1.80300902  & -0.89119131 \\
        $\alpha_1$          & -0.61792703 & -2.91001616 \\
        $A_2$               &             & 0.63210387  \\
        $\alpha_2$          &             & -0.08132401 \\ \hline
        $\Omega_0 T_R$      & 2.958       & 2.936       \\
        $1-F$               & $6\times10^{-11}$ & $4\times10^{-9}$ \\ \hline \hline
    \end{tabular}
    \caption{\textbf{Time-optimal and $\bm{T_R}$-optimal $\bm{\mathrm{CZ}}$ gate parameters.} The table shows the parameters of two pulses implementing a CZ gate in the perfect blockade regime using ansatz~\eqref{eq:ansatz_sin}. The left column corresponds to the time-optimal pulse. The right column describes a pulse that minimizes the time $\Omega_0 T_R$ spent in Rydberg states during gate execution. In the absence of Rydberg-state decay, the gates reach infidelities $1-F$. Both pulse profiles are shown in Fig.~\ref{fig:CZ_pulses}.}
    \label{tab:timeoptimal_CZ}
    \end{table*}

    \begin{table*}[t]
    \begin{tabular}{c|r|r|r|r} \hline \hline
        & Ansatz~\eqref{eq:ansatz_sin_cos} & Ansatz~\eqref{eq:ansatz_sin} & Ansatz~\eqref{eq:ansatz_sin} & Ansatz~\eqref{eq:ansatz_sin} \\
        & 14 parameters & 8 parameters & 6 parameters & 10 parameters \\ \hline
        $\Omega_0 T$        & 10.83264840 & 10.97094681 & 12.24229503 & 12.72911958 \\
        $\Delta_0/\Omega_0$ & -0.00808517 & -0.19566367 & -0.73502207 & -1.43281803 \\
        $A_1$               & 1.29967575  & 0.43131090  & 3.00127734 & -1.98920752 \\
        $\alpha_1$          & -0.15117887 & -1.16460209 & 2.80831494 & -6.31464092 \\
        $B_1$               & 0.26117336  &             &            & \\
        $\beta_1$           & 0.32173635  &             &            & \\
        $A_2$               & -0.18593552 & 1.05669771  & 3.45152105 & 0.10028046 \\
        $\alpha_2$          & -0.37946965 & -0.70545851 & 1.08933841 & 1.08234350 \\
        $B_2$               & -0.33014685 &             &            & \\
        $\beta_2$           & -1.01300629 &             &            & \\
        $A_3$               & 0.07813142  & 0.88054914  &            & 0.71976922 \\
        $\alpha_3$          & -0.48186569 & -0.22756692 &            & -0.43203706 \\
        $B_3$               & 0.36779145  &             &            & \\
        $\beta_3$           & 0.24847061  &             &            & \\
        $A_4$               &             &             &            & 0.65327507 \\
        $\alpha_4$          &             &             &            & 0.20638712 \\ \hline
        $\Omega_0 T_R$      & 4.905       & 4.179       & 4.403      & 3.947 \\
        $1-F$               & $1\times10^{-14}$ & $5\times10^{-9}$ & $2\times10^{-9}$ & $1\times10^{-7}$ \\ \hline \hline
    \end{tabular}
    \caption{\textbf{Time-optimal, minimal-parameter, and $\bm{T_R}$-optimal $\bm{\overline{\mathrm{CCZ}}}$ gate parameters.} The table shows pulse parameters realizing $\overline{\mathrm{CCZ}}$ gates in the perfect blockade regime. The time-optimal pulse~\cite{evered2023highfidelity} requires the general ansatz~\eqref{eq:ansatz_sin_cos} with 14 parameters (left column). A nearly time-optimal gate can be realized with a pulse described by 8 parameters using the symmetric ansatz~\eqref{eq:ansatz_sin}. The minimal number of pulse parameters required to realize the gate is 6. A pulse described by 10 parameters can reduce the Rydberg time $\Omega_0 T_R$ significantly. In the absence of Rydberg-state decay, the gates reach infidelities $1-F$. The pulses are plotted in Fig.~\ref{fig:CCZ_combined}(c).}
    \label{tab:timeoptimal_CCZ}
    \end{table*}

    \begin{table*}[t]
    \begin{tabular}{c|r|r|r|r} \hline \hline
        & Ansatz~\eqref{eq:ansatz_sin_cos} & Ansatz~\eqref{eq:ansatz_sin} & Ansatz~\eqref{eq:ansatz_sin} & Ansatz~\eqref{eq:ansatz_sin} \\
        & 18 parameters & 10 parameters & 8 parameters & 14 parameters \\ \hline
        $\Omega_0 T$        & 11.80271325 & 12.42032121 & 14.14222223 & 15.64882230 \\
        $\Delta_0/\Omega_0$ & 0.13332200  & -0.09371255 & -0.16954382 & -0.62344592  \\
        $A_1$               & 0.66983248  & 1.02359835  & 0.31647908  & -1.21309950 \\
        $\alpha_1$          & 1.31476492  & -0.99945357 & -0.75670150 & 0.30395321  \\
        $B_1$               & 2.81147759  &             &             & \\
        $\beta_1$           & -1.58457206 &             &             & \\
        $A_2$               & -0.36210033 & 1.96361795  & 0.39053196  & 3.65589621  \\
        $\alpha_2$          & -1.57467284 & -0.80000698 & 1.30626162  & 0.07086844  \\
        $B_2$               & -2.17781176 &             &             & \\
        $\beta_2$           & 1.09432979  &             &             & \\
        $A_3$               & 0.41649028  & 0.74691915  & 0.80503678  & -0.90823921 \\
        $\alpha_3$          & -0.64527764 & -0.18123922 & 0.57628361  & -2.60857295 \\
        $B_3$               & 0.47012471  &             &             & \\
        $\beta_3$           & 0.19164886  &             &             & \\
        $A_4$               & 0.14593750  & 0.70514304  &             & -0.12491227 \\
        $\alpha_4$          & 0.51086808  & 0.32656123  &             & 0.98704073  \\
        $B_4$               & 1.08196100  &             &             & \\
        $\beta_4$           & -0.06506773 &             &             & \\
        $A_5$               &             &             &             & 0.31508103  \\
        $\alpha_5$          &             &             &             & 0.34077972  \\
        $B_5$               &             &             &             & \\
        $\beta_5$           &             &             &             & \\
        $A_6$               &             &             &             & -0.35948449 \\
        $\alpha_6$          &             &             &             & -0.80274892 \\ \hline
        $\Omega_0 T_R$      & 5.494       & 4.925       & 6.407       & 4.755       \\
        $1-F$               & $5\times10^{-8}$ & $1\times10^{-8}$ & $2\times10^{-8}$ & $5\times10^{-6}$ \\ \hline \hline
    \end{tabular}
    \caption{\textbf{Time-optimal, minimal-parameter, and $\bm{T_R}$-optimal $\bm{\overline{\mathrm{CCCZ}}}$ gate parameters.} The table shows pulse parameters realizing $\overline{\mathrm{CCCZ}}$ gates in the perfect blockade regime. The time-optimal pulse~\cite{evered2023highfidelity} requires the general ansatz~\eqref{eq:ansatz_sin_cos} with 18 parameters (left column). A nearly time-optimal gate can be realized with a pulse described by 10 parameters using the symmetric ansatz~\eqref{eq:ansatz_sin}. The minimal number of pulse parameters required to realize the gate is 8. A pulse described by 14 parameters can reduce the Rydberg time $\Omega_0 T_R$ significantly. In the absence of Rydberg-state decay, the gates reach infidelities $1-F$. The pulses are plotted in Fig.~\ref{fig:CCCZ_combined}(c).}
    \label{tab:timeoptimal_CCCZ}
    \end{table*}

    \begin{table*}[t]
    \begin{tabular}{c|r|r|r} \hline \hline
        & Ansatz~\eqref{eq:ansatz_sin} & Ansatz~\eqref{eq:ansatz_sin} & Ansatz~\eqref{eq:ansatz_sin} \\
        & 8 parameters & 6 parameters & 8 parameters \\ \hline
        $\Omega_0 T$        & 16.36973127 & 18.17182396 & 16.56456040 \\
        $\Delta_0/\Omega_0$ & -0.28705772 & -0.51050960 & -0.41971218  \\
        $A_1$               & -0.62187299 & -1.03247663 & -0.04643639 \\
        $\alpha_1$          & 2.40051297  & 2.20024245  & -0.83527865 \\
        $A_2$               & 0.28403466  & 0.74044080  & 1.34678303  \\
        $\alpha_2$          & 0.45564291  & -1.34735876 & 0.05257253  \\
        $A_3$               & 1.98974449  &             & 0.81906055  \\
        $\alpha_3$          & -0.20040935 &             & -0.31596455 \\ \hline
        $\Omega_0 T_R$      & 6.541       & 7.867       & 6.105       \\
        $1-F$               & $3\times10^{-8}$ & $6\times10^{-9}$ & $5\times10^{-8}$ \\ \hline \hline
    \end{tabular}
    \caption{\textbf{Time-optimal, minimal-parameter, and $\bm{T_R}$-optimal CZ-CZ-CZ gate parameters.} The table shows pulse parameters realizing CZ-CZ-CZ gates in the perfect blockade regime. The time-optimal pulse requires 8 parameters using ansatz~\eqref{eq:ansatz_sin} (left column). The minimal number of pulse parameters required to realize the gate is 6. A pulse described by 8 parameters can reduce the Rydberg time $\Omega_0 T_R$ significantly. In the absence of Rydberg-state decay, the gates reach infidelities $1-F$. The pulses are plotted in Fig.~\ref{fig:CZCZCZ_combined}(c).}
    \label{tab:timeoptimal_CZCZCZ}
    \end{table*}

\clearpage

\bibliography{refs}

\end{document}